\documentclass[prb,twocolumn,showpacs,amsmath,amssymb,superscriptaddress]{revtex4}

\usepackage{graphicx}

\begin{document}

\title{Real-time renormalization group and cutoff scales in
  nonequilibrium applied to an arbitrary quantum dot in the Coulomb
  blockade regime}
\author{Thomas Korb}
\author{Frank Reininghaus}
\author{Herbert Schoeller}
\affiliation{Institut f\"ur Theoretische Physik, Lehrstuhl A, RWTH Aachen, 52056 Aachen, Germany}
\author{J\"urgen K\"onig}
\affiliation{Institut f\"ur Theoretische Physik III, Ruhr-Universit\"at Bochum, 44780 Bochum, Germany}
\date{\today}
\begin{abstract}
We apply the real-time renormalization group (RG) in nonequilibrium to
an arbitrary quantum dot in the Coulomb blockade regime. Within
one-loop RG-equations, we include self-consistently the kernel
governing the dynamics of the reduced density matrix of the dot. As a
result, we find that relaxation and dephasing rates generically cut off
the RG flow. In addition, we include all other cutoff scales defined by
temperature, energy excitations, frequency, and voltage. We apply the
formalism to transport through single molecular magnets, realized by
the fully anisotropic Kondo model (with three different exchange
couplings $J_x$, $J_y$, and $J_z$) in a magnetic field $h_z$. We
calculate the differential conductance as function of bias voltage $V$
and discuss a quantum phase transition which can be tuned by changing
the sign of $J_x J_y J_z$ via the anisotropy parameters. Finally, we
calculate the noise $S(\Omega)$ at finite frequency $\Omega$ for the
isotropic Kondo model and find that the dephasing rate determines the
height of the shoulders in $dS(\Omega)/d\Omega$ near $\Omega=V$.
\end{abstract}

\maketitle

\section{Introduction}
\label{sec:introduction}

A fundamental issue of recent interest is the development of
renormalization group (RG) methods in
nonequilibrium.\cite{rtrg,rtrg_dot,rtrg_spinboson,coleman_etal,rosch_etal,millis,kehrein,jakobs_meden_hs,gezzi_etal}
Besides the discovery of new power law exponents for the conductance
induced by nonequilibrium occupation
probabilities,\cite{jakobs_meden_hs} an interesting question was
raised whether voltage-induced decay rates provide additional cutoffs
of the RG flow.\cite{kaminski_etal,rosch_etal} In this context, the
nonequilibrium Kondo model has been discussed, which can be realized
by a single-level quantum dot (QD) in the Coulomb-blockade (CB) regime
coupled via spin exchange processes $J_{\alpha\alpha'}$ to two
reservoirs $\alpha=L,R$. In the isotropic case and above all cutoff
scales, the exchange couplings $J=J_{\alpha\alpha'}$ are all the same
and are enhanced by reducing the band width $\Lambda$ of the
reservoirs according to the poor man's scaling equation
$\frac{dJ}{dl}=2J^2$, \cite{poor_man_scaling} with
$l=\ln(\Lambda_0/\Lambda)$ ($\Lambda_0$ denotes the inital band
width). The enhanced screening of the dot spin leads to the Kondo
effect with unitary conductance below the Kondo temperature
$T_K=\Lambda_0\exp[-1/(2J)]$ (Ref.~\onlinecite{kondo_theo}) (for
experiments in quantum dots see, e.g.,
Refs.~\onlinecite{kondo_exp}). However, for voltages $V\gg T_K$, it
was argued that the system cannot reach the strong coupling fixed
point since the nondiagonal coupling constants, $J_{LR}=J_{RL}$, are
cut off by the voltage \cite{coleman_etal,rosch_etal} and the diagonal
ones, $J_{LL}$ and $J_{RR}$, by the voltage-induced decay rate $\Gamma
= \pi J_{LR}^2|_{\Lambda=V} V$.\cite{rosch_etal,kehrein} This has
raised the fundamental question how decay processes can be implemented
in nonequilibrium RG. Applying flow equation methods to the isotropic
Kondo model without magnetic field, it was shown within a two-loop
formalism in Ref.~\onlinecite{kehrein} that the inclusion of a
third-order term $\sim J^3$ in the RG equation leads to a cutoff of
the RG flow at the scale $\Gamma$. 

In this paper, we will analyze this problem from a more general point of view and will
show within a microscopic one-loop RG formalism that relaxation and dephasing rates will
always cut off the RG flow for an arbitrary quantum dot in the Coulomb blockade regime.
This confirms the conjecture of Refs.~\onlinecite{rosch_etal} and~\onlinecite{kaminski_etal} and generalizes the analysis 
of Ref.~\onlinecite{kehrein} to an arbitrary QD (including orbital and spin fluctuations, many levels,
interference effects, etc.). We propose to use 
the real-time RG (RTRG) formalism of Ref.~\onlinecite{rtrg} with a cutoff defined in
frequency space since this approach directly discusses the time evolution of
the reduced density matrix of the dot via a kinetic equation. Within this
formalism, the decay rates occur naturally as the negative imaginary parts of the  
eigenvalues of the kernel determining the dissipative part of the kinetic equation. This
has already been demonstrated previously by applying RTRG to the calculation of 
steady-state transport through quantum dots in the charge fluctuation regime \cite{rtrg_dot} 
and to the study of the real-time evolution of the occupation probabilities within 
the spin boson model.\cite{rtrg_spinboson} Another advantage of the RTRG approach is
the fact that the kernel can easily be inserted self-consistently into the one-loop RG equations
of the coupling parameters (analogous to self-energy insertions within Green's function
techniques), providing the unique possibility to obtain the physical decay rates
within a nonequilibrium one-loop RG formalism. In addition, we also provide a microscopic
formalism from which all other standard cutoff scales, such as temperature, energy excitations
(e.g., magnetic fields), frequencies, and voltages, can be deduced analytically. 

The original RTRG \cite{rtrg} was formulated
with a cutoff defined in time space for the reservoir correlation function. This makes
it technically difficult to apply the formalism to problems where the interaction between dot and
reservoirs is nonlinear as it is the case for quantum dots in the cotunneling regime,
where orbital and spin fluctuations dominate transport. Therefore, we use in this work
a cutoff defined in frequency space but adapt the same formalism to set up the RG equations
as in Ref.~\onlinecite{rtrg}. This leads to a combined time-frequency formalism since the
time-ordering of the renormalized vertices is needed due to their operator nature (the
degrees of freedom of the dot are not integrated out within RTRG, and therefore all
coupling vertices are operators acting on the dot degree of freedom). The only disadvantage
of the analytic formalism presented in this work is still the fact that the irrelevant prefactors of 
the various decay rates cutting off the RG flow cannot be determined unambigiously; this has to be
left for future developments.

First, we apply the formalism to quantum transport through single molecular magnets (SMM). Recently,
it has been shown that the study of Kondo physics can be used for
transport spectroscopy of SMM,\cite{romeike_etal,SMM_strong} i.e., the
various anisotropy parameters determining the spin 
excitation spectrum can be identified. In the regime where the Kondo temperature is smaller
than the distance to the next spin excitation, it has been shown that a pseudo-spin-1/2 model
can be derived which can be mapped onto the fully anisotropic Kondo model with three different
exchange couplings $J_x$, $J_y$, and $J_z$. Interestingly, this model reveals a quantum 
phase transition by changing the sign of $J_x J_y J_z$, separating the
flow to the weak and strong
coupling regimes. Since the exchange couplings depend on the transverse anisotropy parameters, which
in turn depend on the coupling of the SMM to the leads, this phase transition can be tuned
in an experimental setup. Using RTRG, we calculate the differential conductance $G(V)$ as function of
bias voltage at finite magnetic field and show that the Kondo-enhanced conductance
at $V=h$, where $h$ is the level spacing between the ground state and
the first excited state, disappears by tuning the system through the phase transition.

Second, we calculate the quantum noise $S(\Omega)$ as function of frequency for
the isotropic Kondo model at finite bias and zero magnetic field. We find that
the dephasing rate can be identified by studying the derivative of the noise 
near $\Omega=V$. Specifically, it turns out that the noise has a dip at $V=\Omega$ (see also
Ref.~\onlinecite{kondo_noise}, where the noise has been calculated for the Toulouse point), whereas
the derivative $dS(\Omega)/d\Omega$ shows a characteristic shoulder with a height depending
on the dephasing rate.

The paper is organized as follows: In Sec.~\ref{sec:model}, we set up the general model and
show the relation to the nonequilibrium Kondo model. Section \ref{sec:perturbation_series}
summarizes the diagrammatic language in Liouville space. Section \ref{sec:rg_formalism} is
the central technical part where we set up the RG equations and explain how
decay rates cut off the RG flow. Finally, we apply the formalism in Sec.
\ref{sec:smm} to transport through single molecular magnets and in
Sec.~\ref{sec:noise} to the calculation of quantum noise. Two appendices 
provide further details of the RG formalism.

\section{Model}
\label{sec:model}

We consider an arbitrary quantum dot coupled to reservoirs via tunneling
processes,
\begin{equation}
H = H_{\text{res}} + H_\text{D} + H_T,
\end{equation}
where $H_{\text{res}}$, $H_\text{D}$, and $H_T$ denote the Hamiltonians
of the reservoirs, the dot, and the tunneling, respectively.
\begin{equation}
H_{\text{res}} = \sum_\alpha H_{\text{res}}^\alpha =
\sum_{k\alpha\sigma} \epsilon_{k\alpha\sigma} 
a^\dagger_{k\alpha\sigma}a_{k\alpha\sigma}
\end{equation}
describes the noninteracting Hamiltonian of the reservoirs with
$a^\dagger_{k\alpha \sigma}$ ($a_{k\alpha \sigma}$) the 
creation (annihilation) operators. $\alpha$ is the reservoir index,
$\sigma$ denotes the spin, and $k$ is an index for the 
single-particle states in the reservoirs. Each reservoir is
assumed to be infinitely large and described by a grand canonical
distribution with electrochemical potential $\mu_\alpha$ and temperature $T$.
The isolated dot Hamiltonian is written in diagonalized form as
\begin{equation}
H_\text{D} = \sum_s E_s |s\rangle\langle s|,
\end{equation}
where $s$ is an index for the many-body eigenstates of the dot with
energy eigenvalues $E_s$. Finally, the interaction between dot and reservoirs is 
described by the standard tunneling Hamiltonian
\begin{equation}
H_T = \sum_{\alpha k l \sigma}t^{\alpha\sigma}_{kl}
a^\dagger_{k\alpha \sigma}c_{l\sigma}\,\,+\,\,\text{H.c.},
\end{equation}
where $c_{l\sigma}$ annihilates a particle with spin $\sigma$
in the single-particle level $l$ on the dot and 
$t^{\alpha\sigma}_{kl}$ denotes the tunneling matrix element.

Since the reservoirs are infinitely large, we describe their spectrum
by the continuum density of states 
$\rho_\mu(\omega)=\sum_k\delta(\omega-\epsilon_{k\mu}+\mu_\alpha)$, with 
$\mu\equiv \alpha\sigma$ an index containing the reservoir and
the spin index (this will be used implicitly in the following). For
the general discussion, we include the case of spin- and 
frequency-dependent density of states in the reservoirs. 
We introduce the continuum fields
\begin{equation}
a_{\mu+}(\omega)={\frac{1}{\sqrt{\rho_\mu(\omega)}}}
\sum_k \delta(\omega-\epsilon_{k\mu}+\mu_\alpha) a^\dagger_{k\mu}
\end{equation}
and $a_{\mu-}(\omega)=a_{\mu+}(\omega)^\dagger$ which fulfill the 
anticommutation relation 
$\{a_{\mu\eta}(\omega),a_{\mu'\eta'}(\omega')\}=
\delta_{\eta,-\eta'}\delta_{\mu\mu'}\delta(\omega-\omega')$. 
With this notation, the
reservoir Hamiltonian and the tunneling part can be written as
\begin{eqnarray}
H_{\text{res}} &=& 
\sum_{\mu}\int d\omega \,(\omega+\mu_\alpha) a_{\mu+}(\omega)
a_{\mu-}(\omega), \label{h_res}\\
H_T &=&
\sum_\mu\int d\omega \, a_{\mu+}(\omega)g_\mu(\omega)
\,\,+\,\,\text{H.c.}, \label{h_T}
\end{eqnarray}
with
\begin{equation}
g_\mu(\omega)=\sqrt{\rho_\mu(\omega)}\,\sum_l
t^\mu_l(\omega) c_{l\sigma},
\end{equation}
where $t^{\alpha\sigma}_l(\omega)\equiv t^{\alpha\sigma}_{kl}$ is the 
tunneling matrix element
in the continuum notation evaluated for reservoir state $k$ such
that $\epsilon_{k\alpha\sigma}-\mu_\alpha=\omega$. 
The contraction of two reservoir field
operators with respect to the equilibrium reservoir distribution 
is given by
\begin{equation}
\label{contraction}
\langle a_{\mu\eta}(\omega)a_{\mu'\eta'}(\omega')\rangle_{\text{res}}=
\delta_{\eta,-\eta'}\delta_{\mu\mu'}\delta(\omega-\omega')\theta_\omega
f^\eta_\omega,
\end{equation}
where $f^+_\omega=f_\omega$ and $f^-_\omega=1-f_\omega=f_{-\omega}$,
with $f_\omega=1/[\exp(\beta\omega)+1]$ denoting the Fermi function
(note that the different electrochemical potentials of the reservoirs
occur in our notation via the interaction picture from the time-dependence
of the field operators). $\theta_\omega = \theta(\Lambda_0-\omega)$
contains the initial band width $\Lambda_0$ of the reservoirs (which are
assumed to be all the same relative to the corresponding electrochemical
potentials).

We now consider a quantum dot in the Coulomb blockade regime, i.e.,
the total charge is fixed, and only cotunneling processes via
virtual intermediate states can lead to orbital and spin fluctuations.
The effective Hamiltonian in this regime is standardly derived using
the Schrieffer-Wolff transformation,\cite{schrieffer_wolff} leading
to
\begin{equation}
\label{h_eff}
H_{\text{eff}} = H_{\text{res}} + H_\text{D} + V_{\text{eff}},
\end{equation}
with
\begin{eqnarray}
\label{v_eff}
V_{\text{eff}} &=& \sum_{\mu\mu'}\int^{\Lambda_0}_{-\Lambda_0}
d\omega d\omega'\,
\left\{g^+_{\mu\mu'}(\omega,\omega')\,
a_{\mu+}(\omega)a_{\mu'-}(\omega')\right.\nonumber\\
&& \hspace{1cm}-\,\left.g^-_{\mu\mu'}(\omega,\omega')\,
a_{\mu'-}(\omega')a_{\mu+}(\omega)\right\},
\end{eqnarray}
where
\begin{eqnarray}
g^+_{\mu\mu'}(\omega,\omega')&=& \frac12\sum_{ss'}
|s\rangle\langle s'| \cdot\langle s|g_{\mu}(\omega)\\
&&\hspace{-2.5cm}
\left(\frac1{\omega+\mu_{\alpha}+E_{s}-H_\text{D}}+
\frac1{\omega'+\mu_{\alpha'}+E_{s'}-H_\text{D}}\right)
g_{\mu'}(\omega')^\dagger|s'\rangle \nonumber
\end{eqnarray}
corresponds to virtual processes, where the electron first
hops from the reservoir to the dot and then back, and
\begin{eqnarray}
g^-_{\mu\mu'}(\omega,\omega')&=& \frac12\sum_{ss'}
|s\rangle\langle s'|\cdot\langle s|g_{\mu'}(\omega')^\dagger \\
&&\hspace{-2.5cm}
\left(\frac1{\omega+\mu_{\alpha}-E_{s'}+H_\text{D}}+
\frac1{\omega'+\mu_{\alpha'}-E_{s}+H_\text{D}}\right)
g_{\mu}(\omega)|s'\rangle \nonumber
\end{eqnarray}
describes the reverse process. Processes where two electrons hop on
or off the dot are not written here but can easily be incorporated
(they are only important for molecular systems with negative 
Coulomb interaction; see Ref.~\onlinecite{vonoppen}).

In normal-ordered form (with respect to the equilibrium 
reservoir distribution), denoted by the symbol $:\dots:$,
we get from Eqs~(\ref{h_eff}), (\ref{v_eff}), and~(\ref{contraction})
\begin{equation}
\label{h_final}
H_{\text{eff}} = H_{\text{res}} + H_\text{D}^{\text{eff}} +
:V_{\text{eff}}:,
\end{equation}
with a renormalized dot Hamiltonian
\begin{equation}
H_\text{D}^{\text{eff}} = H_\text{D} + \sum_{\mu\eta} \int d\omega \,\theta_\omega
\,\eta \,g^\eta_{\mu\mu}(\omega,\omega)\,f^\eta_\omega,
\end{equation}
which can contain logarithmic energy renormalizations of the dot states
due to orbital interferences or due to spin-dependent tunneling matrix
elements (see, e.g., Refs.~\onlinecite{boese_etal} and~\onlinecite{koenig_braun}) (for the
Kondo model under consideration in this work, such renormalizations do
not occur). Finally, the normal-ordered interaction term reads
\begin{multline}
V\equiv :V_{\text{eff}}:{}=
\sum_{\mu\mu'}\int_{-\Lambda_0}^{\Lambda_0}d\omega d\omega'\\
g_{\mu\mu'}(\omega,\omega'):a_{\mu+}(\omega)a_{\mu'-}(\omega'):,
\label{v_final}
\end{multline}
with
\begin{equation}
g_{\mu\mu'}(\omega,\omega')=\sum_\eta g^\eta_{\mu\mu'}(\omega,\omega').
\end{equation}

Equations~(\ref{h_final}) and~(\ref{v_final}) are the final general form of the 
model under consideration, which is the starting point for the 
renormalization group formalism. It is still completely general,
except for the fact that the dot is assumed to be at fixed charge.
In order to simplify the notation, we omit in the following the
index \lq\lq eff\lq\lq, and use the short-hand notation
\begin{equation}
\label{v_short}
V = g_{11'}:a_{1+} a_{1'-}:,
\end{equation}
where we sum/integrate implicitly over 
the indices $1\equiv\omega_1\mu_1$ and $1'\equiv\omega_{1'}\mu_{1'}$. 
We note the property
\begin{equation}
\label{g_property}
g_{11'} = g_{1'1}^*,
\end{equation}
which guarantees the Hermiticity of $H_\text{D}$.

The fully anisotropic Kondo model
under consideration in Sec.~\ref{sec:smm} is realized for
the special case where the dot Hamiltonian consists of two states with
(pseudo-) spin up or down (i.e., $s=\pm=\uparrow,\downarrow$ denotes the
dot spin). The antiferromagnetic exchange processes between the dot spin 
$\underline{S}$ and the reservoir spins are described by the coupling
\begin{equation}
\label{g_kondo} 
g_{11'}=\frac12\sum_{i=x,y,z}
J^i_{\alpha_1\alpha_1^\prime}S^i\,\sigma^i_{\sigma_1\sigma_1^\prime},
\end{equation}
where $\sigma^i$, $i=x,y,z$, are the Pauli matrices. Inserting this
into Eq.~(\ref{v_final}) gives the standard form of the anisotropic
Kondo model,
\begin{eqnarray}
\label{H_D_kondo}
H_\text{D}&=& \frac{h}{2}\sum_{s=\pm}s|s\rangle\langle s|,\\
\label{v_kondo}
V&=&\frac12\sum_{i\mu\mu'}\int^{\Lambda_0}_{-\Lambda_0} d\omega d\omega' \,
J^i_{\alpha\alpha'}\, S^i\,
\sigma^i_{\sigma\sigma'}\\\nonumber &&\hspace{1.5cm}
:a_{\mu+}(\omega) a_{\mu'-}(\omega'):,
\end{eqnarray}
where $h$ denotes the effective magnetic field in the $z$ direction.
In contrast to the usual case $J^x=J^y$, we discuss here the fully anisotropic
Kondo model with three different exchange couplings, a model of recent interest
if the dot is replaced by a single molecular magnet.\cite{romeike_etal} In this
case, the isolated molecule is described by the spin Hamiltonian 
\begin{equation}
\label{h_mol}
H_{\text{mol}}=-D(S_\text{M}^z)^2-\frac12\sum_n B_{2n}\left[(S_\text{M}^+)^{2n}+
(S_\text{M}^-)^{2n}\right]+h_z S_\text{M}^z,
\end{equation} 
where $D$ and $B_{2n}$ denote the longitudinal and transverse anisotropy 
constants, and the original spin $S_\text{M}$ is greater than $1/2$.
$h_z$ denotes the physical magnetic field in the $z$ direction.
If one projects an isotropic exchange 
$(J/2)\underline{S}_M \underline{\sigma}_{\sigma\sigma'}
a_{\mu +}(\omega)a_{\mu' -}(\omega')$
between the molecule and reservoirs 
onto the two lowest eigenstates $|\pm\rangle$ of $H_{\text{mol}}$ (which is
justified when the Kondo temperature is lower than the first magnetic
excitation), one obtains a pseudo-spin-$1/2$ model described by the
fully anisotropic Kondo model [Eqs.~(\ref{H_D_kondo}) and~(\ref{v_kondo})] with  
\begin{eqnarray}
\label{magnetic_field}
h &=& \langle +|H_\text{mol}|+\rangle - \langle -|H_\text{mol}|-\rangle,\\
\label{xy_exchange_couplings}
J^{x/y} &=& J \langle +|S_\text{M}^+ \pm S_\text{M}^-|-\rangle \label{jxy},\\ 
\label{z_exchange_couplings}
J^z     &=& 2J\langle +|S_\text{M}^z|+\rangle>0 \label{jz},
\end{eqnarray}
where $J$ is the isotropic exchange constant between the original molecular
spin and the reservoirs (see Ref.~\onlinecite{romeike_etal} for further details).

\section{Perturbation series}
\label{sec:perturbation_series}

We aim at calculating the stationary dot distribution $p^{\text{st}}$, the
stationary current $I^\gamma_{\text{st}}$ in
lead $\gamma$, and the frequency-dependent noise power 
\begin{eqnarray}
S^{\gamma\gamma'}_\Omega&=& \frac12\int dt e^{i\Omega t}
\langle\{\delta I^\gamma(t),\delta I^{\gamma'}(0)\}\rangle \nonumber\\ 
&=&\bar{S}^{\gamma\gamma'}_{-\Omega}+\bar{S}^{\gamma'\gamma}_\Omega
-2\pi\delta(\Omega)I^\gamma_\text{st}I^{\gamma'}_\text{st},
\label{noise1}
\end{eqnarray}
with $\delta I^\gamma = I^\gamma - I^\gamma_\text{st}$ and
\begin{equation}
\label{noise2}
\bar{S}^{\gamma\gamma'}_\Omega=\frac12\int_{-\infty}^0 dt e^{-i\Omega t}
\langle\{I^\gamma(t),I^{\gamma'}(0)\}\rangle.
\end{equation}
Due to current conservation, we have 
\begin{equation}
\label{noise_symmetry}
\sum_\gamma S^{\gamma\gamma'}_\Omega =
\sum_{\gamma'} S^{\gamma\gamma'}_\Omega = 0,
\end{equation} 
and therefore, for two reservoirs, it is sufficient to calculate
the diagonal noise $S^{\gamma\gamma}$.

The current operator $I^\gamma$ for lead $\gamma$ is given by 
$I^\gamma=-(d/dt)N_\gamma=-i[V,N_\gamma]$, where $N_\gamma$ is the
particle number in reservoir $\gamma$ (we use units $e=\hbar=1$). 
Using Eq.~(\ref{v_short}), this gives
\begin{equation}
I^\gamma = i(\delta_{\gamma\alpha_1}-\delta_{\gamma\alpha_1^\prime})
g_{11'}:a_{1+} a_{1'-}:.
\end{equation}
Following Ref.~\onlinecite{rtrg}, we start from an initial distribution 
$\rho(t_0)=p(t_0)\rho_{\text{res}}$ which factorizes 
into an arbitrary dot part $p(t_0)$ and an equilibrium grand canonical
distribution $\rho_{\text{res}}=\Pi_\alpha 
\exp[-\beta(H_{\text{res}}^\alpha-\mu_\alpha N_\alpha)]/Z_\alpha$ for
the reservoirs. The reduced density matrix of the dot at time $t$ can
then be written as
\begin{equation}
\label{p(t)}
p(t) = \text{Tr}_{\text{res}} e^{-iL(t-t_0)}p(t_0)\rho_{\text{res}},
\end{equation}
where $L=[H,\cdot]$ is the Liouville operator, which is a superoperator
acting on ordinary operators $b$ via $Lb=[H,b]$. According to Eq.~(\ref{h_final}),
we decompose $L=L_{\text{res}}+L_\text{D}+L_V$. Using Eq.~(\ref{v_short}), the
Liouville operator of the interaction part can be written as
\begin{equation}
\label{L_V}
L_V=[V,\cdot]=p^\prime G^{pp'}_{11'}:J^p_{1+}J^{p'}_{1'-}:, 
\end{equation}
where we sum implicitly over the Keldysh indices $p,p'=\pm$. Here, 
\begin{equation}
\label{G_vertex}
G^{pp'}_{11'}=\delta_{pp'}G^{pp}_{11'} 
\end{equation}
and $J^p_{1\eta}$ are
superoperators acting on usual dot (lead) operators $b$ via
\begin{align}
\label{G_initial}
G^{++}_{11'}b&=g_{11'}b, & G^{--}_{11'}b&=-bg_{11'},\\ 
J^+_{1\eta}b&=a_{1\eta}b, & J^-_{1\eta}b&=ba_{1\eta}.
\end{align}
Taking matrix elements with respect to the dot states, the superoperators
$L_\text{D}$ and $G^{pp}_{11'}$ are given by
\begin{eqnarray}
\label{L_matrix}
\hspace{-0.5cm}
(L_\text{D})_{s_1 s_1^\prime,s_2 s_2^\prime} &=&
(H_\text{D})_{s_1 s_2}\delta_{s_1^\prime s_2^\prime}-
\delta_{s_1 s_2}(H_\text{D})_{s_2^\prime s_1^\prime},\\
\label{G+_matrix}
\hspace{-0.5cm}
(G^{++}_{11'})_{s_1 s_1^\prime,s_2 s_2^\prime} &=&
(g_{11'})_{s_1 s_2}\delta_{s_1^\prime s_2^\prime},\\
\label{G-_matrix}
\hspace{-0.5cm}
(G^{--}_{11'})_{s_1 s_1^\prime,s_2 s_2^\prime} &=&
-\delta_{s_1 s_2}(g_{11'})_{s_2^\prime s_1^\prime}.
\end{eqnarray}
If the states $|s\rangle$ are the eigenstates of $H_\text{D}$
with eigenvalues $E_s$, we get
\begin{equation}
\label{L_matrix_diagonal}
(L_\text{D})_{s_1 s_1^\prime,s_2 s_2^\prime} =
(E_{s_1}-E_{s_1^\prime})\delta_{s_1 s_2}
\delta_{s_1^\prime s_2^\prime}.
\end{equation}
From these matrix representations, we get 
\begin{equation}
\label{LG_property}
\sum_s (L_\text{D})_{ss,\cdot\cdot}=0,\quad\quad
\sum_p\sum_s (G^{pp}_{11'})_{ss,\cdot\cdot}=0,
\end{equation}
which is an important property guaranteeing the conservation of
probability $\sum_s p(t)_{ss}=1$ (see Ref.~\onlinecite{rtrg}).

Following Ref.~\onlinecite{rtrg}, we expand Eq.~(\ref{p(t)}) in $L_V$ and integrate 
out the leads in order to get an effective description for the dynamics of the dot. We 
define the interaction picture of $L_V$ with respect to $L_{\text{res}}+L_\text{D}$
and obtain $L_V(t)=p^\prime G^{pp'}_{11',t}:J^p_{1+}J^{p'}_{1'-}:$, with 
\begin{equation}
\label{G_intpic}
G^{pp'}_{11',t}=e^{i(\omega_1-\omega_1^\prime+\mu_{\alpha_1}-\mu_{\alpha_1^\prime})t}\,
e^{iL_{\text{D}}t}G^{pp'}_{11'}e^{-iL_{\text{D}}t}.
\end{equation}
Each term in the perturbation expansion is then averaged over the 
equilibrium reservoir distribution by using Wick's theorem. Using
Eq.~(\ref{contraction}), this leads to pair contractions between the
superoperators $J^p_{1\eta}$ given by
\begin{equation}
\label{super_contraction}
\gamma_{1\eta,1'\eta'}^{pp'}=
{J^p_{1\eta} J^{p'}_{1'\eta'}
  \begin{picture}(-20,11) 
    \put(-30,8){\line(0,1){3}} 
    \put(-30,11){\line(1,0){15}} 
    \put(-15,8){\line(0,1){3}}
  \end{picture}
  \begin{picture}(20,11) 
  \end{picture}}
=\langle J^p_{1\eta} J^{p'}_{1',\eta'}\rangle =\delta_{11'}\delta_{\eta,-\eta'}
f^{p'\eta}_{\omega_1}\theta_{\omega_1}.
\end{equation}
In this way, we obtain a sequence of time-ordered dot superoperators 
$(-i)G^{pp'}_{11',t}$ in interaction picture, 
connected in an arbitrary way by lead contractions (for details
and diagrammatic representations on the time axis, see Ref.~\onlinecite{rtrg}). 

The series of all diagrams can be grouped in irreducible and reducible parts,
where irreducible means that any vertical cut to the time axis hits at least
one reservoir contraction (see Fig.~\ref{diagram_example} for an example).
We define the kernel $\Sigma_\Omega=\int^\infty_0 dt\,
e^{i\Omega t} \Sigma(t)$ in Laplace space, where $\Sigma(t)$ is the sum 
of all irreducible diagrams between time $0$ and $t$. The whole series
of all diagrams can then be formally resummed and we obtain the
following result for the dot distribution in Laplace space:

\begin{figure}
\includegraphics[scale=1]{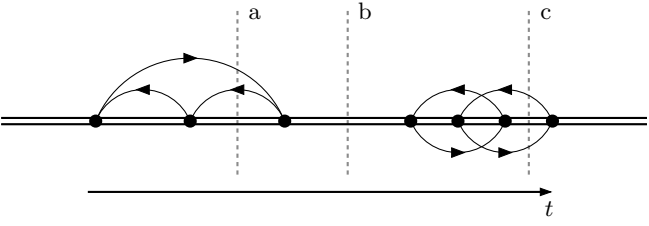}
\caption{Example of a sequence of two irreducible blocks. The double line represents
time propagation in Liouville space of the dot (time increases to the right). 
The black dots represent the 
vertices $G^{pp'}_{11'}$. The lines connecting the vertices are the reservoir
contractions. Whereas the auxiliary vertical lines $a$ and $c$ hit reservoir
contractions, line $b$ does not and separates the two irreducible blocks.}
\label{diagram_example}
\end{figure}

\begin{equation}
\label{distribution_omega}
p_\Omega = \Pi_\Omega \,p(t_0),\quad\quad \Pi_\Omega=\frac{i}{\Omega-L_\text{D}-i\Sigma_\Omega},
\end{equation}
with $p_\Omega=\int_{t_0}^\infty dt \,e^{i\Omega t}p(t)$. The stationary 
distribution follows from $p^{\text{st}}=-i\lim_{\Omega\rightarrow 0}\Omega \,p_\Omega$, 
leading to
\begin{equation}
\label{distribution_stationary}
(L_\text{D} + i\Sigma)\,p^{\text{st}}=0,
\end{equation}
where $\Sigma=\Sigma_{\Omega=0}$.
Therefore, the central quantity to be calculated within renormalization
group is the irreducible kernel $\Sigma$; the stationary distribution then follows from
finding the eigenvector with eigenvalue zero of $L_\text{D}+i\Sigma$. We note that the
irreducible kernel starts and ends with two boundary vertices, denoted by $B$ and
$A$, respectively. Compared to Eq.~(\ref{G_intpic}), their interaction picture is 
slightly differently defined and contains the frequency $\Omega$,
\begin{eqnarray}
\label{A_intpic}
A^{pp'}_{11'\Omega,t}&=&e^{i\Omega t}e^{i(\omega_1-\omega_1^\prime+\mu_{\alpha_1}-\mu_{\alpha_1^\prime})t}\,
A^{pp'}_{11'\Omega}e^{-iL_{\text{D}}t},\\
\label{B_intpic}
B^{pp'}_{11'\Omega,t}&=&e^{-i\Omega t}e^{i(\omega_1-\omega_1^\prime+\mu_{\alpha_1}-\mu_{\alpha_1^\prime})t}\,
e^{iL_{\text{D}}t}B^{pp'}_{11'\Omega}.
\end{eqnarray}
Before starting the RG, we have $A^{pp'}_{11'\Omega}=B^{pp'}_{11'\Omega}=G^{pp'}_{11'}$, but during RG,
the boundary vertices renormalize differently and can become $\Omega$ dependent.

A similiar approach can be set up for the calculation of current and noise.
Choosing $t_0=0$, we write for the current in Laplace space
\begin{equation}
I^\gamma_\Omega=\int_0^\infty e^{i\Omega t}\,\text{Tr}
\,T[-iL_V^\gamma(t)]e^{-i\int_0^t dt' L_V(t')}p(0)\rho_{\text{res}},
\end{equation}
where $T$ is the time-ordering symbol and 
\begin{equation}
L_V^\gamma=\frac12i\{I^\gamma,\cdot\}=
p^\prime G^{\gamma,pp'}_{11'}:J^p_{1+}J^{p'}_{1'-}:
\end{equation} 
is the current superoperator. The current vertex in
Liouville space is given by
\begin{equation}
\label{I_vertex}
G^{\gamma,pp'}_{11'}=c^\gamma_{\alpha_1\alpha_1^\prime}p'
G^{pp'}_{11'},
\end{equation}
with
\begin{equation}
c^\gamma_{\alpha\alpha'}=-\frac12(\delta_{\gamma\alpha}-
\delta_{\gamma\alpha'}).
\end{equation}
Expanding the
exponential as described above in $L_V$, integrating out the reservoirs, and 
resumming the whole series using the irreducible blocks, one arrives at
\begin{equation}
\label{current_omega}
I^\gamma_\Omega=\text{Tr}_\text{D}\Sigma^\gamma_\Omega \,p_\Omega,
\end{equation}
and the stationary current follows from
\begin{equation}
\label{current_stationary}
I^\gamma_\text{st}=\text{Tr}_\text{D}\Sigma^\gamma \,p^\text{st},
\end{equation}
with $\Sigma^\gamma=\Sigma^\gamma_{\Omega=0}$.
Here, $\text{Tr}_{\text{D}}$ denotes the trace over the dot states and
$\Sigma^\gamma_\Omega$ is the irreducible kernel containing exactly one 
current vertex $G^{\gamma,pp'}_{11'}$ with interaction picture defined by
\begin{equation}
\label{I_intpic}
G^{\gamma,pp'}_{11'\Omega,t}=e^{i\Omega t}e^{i(\omega_1-\omega_1^\prime+\mu_{\alpha_1}-\mu_{\alpha_1^\prime})t}\,
e^{iL_{\text{D}}t}G^{\gamma,pp'}_{11'\Omega}e^{-iL_{\text{D}}t}.\\
\end{equation}

For the noise, we choose $t_0=-\infty$ and start from the expression
\begin{multline}
\bar{S}^{\gamma\gamma'}_\Omega=
\int_{-\infty}^0 dt\,e^{-i\Omega t}\,\text{Tr}\,
T\,[-iL_V^\gamma(0)][-iL_V^{\gamma'}(t)]\\ \times
e^{-i\int_{-\infty}^0 dt' L_V(t')}p(-\infty)\rho_\text{res}.
\end{multline}
Again, expanding in $L_V$, integrating out the reservoirs, and resumming
via irreducible blocks gives 
\begin{equation}
\label{noise_omega}
\bar{S}^{\gamma\gamma'}_\Omega=\text{Tr}_{\text{D}}
(\Sigma^{\gamma\gamma'}_\Omega
+\Sigma^\gamma_\Omega\Pi_\Omega{\Sigma^{\gamma'}_\Omega}^\dagger)\,
p^{\text{st}},
\end{equation}
with $\Sigma^{\gamma\gamma'}_\Omega$ the
irreducible kernel containing exactly two current vertices (see also 
Ref.~\onlinecite{noise}). Since the current vertices can also lie at the two 
boundaries of the kernel, one has to define several boundary current 
vertices with slightly different interaction picture
compared to Eq.~(\ref{I_intpic}) (see Appendix B for more details).

\section{Renormalization Group formalism and cutoff scales}
\label{sec:rg_formalism}

We now take a reduced band width $\Lambda$ in the definition of the contraction
[Eq.~(\ref{super_contraction})] by replacing $\theta_\omega\rightarrow 
\theta(\Lambda-|\omega|)$.
Following Ref.~\onlinecite{rtrg}, we determine the $\Lambda$-dependence of 
$L_{\text{D}}$ and $G_{11'}$ in such a way that the total sum of all diagrams 
remains invariant. This leads to the RG diagrams of 
Fig.~\ref{rg_diagrams} which are evaluated in Appendix A with the result
\begin{eqnarray}
\left(\frac{dG^{p_1p_1^\prime}_{11'}}{d\Lambda}\right)_{ik} &=&
i\int_0^{\Gamma_j^{-1}} \hspace{-0.4cm}dt 
\,\delta_{\omega_2}\nonumber\\
&&\hspace{-2cm}\times\left\{p_2^\prime f^{-p_2^\prime}_{\omega_2}\, 
(G^{p_1p_2}_{12,t/2})_{ij}\,(G^{p_2^\prime p_1^\prime}_{21',-t/2})_{jk}
\,\right.\nonumber\\
&&\hspace{-1cm}\left. -p_2 f^{p_2}_{\omega_2}\,
(G^{p_2^\prime p_1^\prime}_{21',t/2})_{ij}\,(G^{p_1 p_2}_{12,-t/2})_{jk}
\right\},\label{rg_G}\\
\left(\frac{dL_{\text{D}}}{d\Lambda}\right)_{ik} &=&
i\int_0^{\Gamma_j^{-1}} \hspace{-0.4cm} dt 
\,\frac{d}{d\Lambda}(\theta_{\omega_1}\theta_{\omega_2})
p_2 p_2^\prime \, f^{p_2^\prime}_{\omega_1}\,f^{-p_2}_{\omega_2}
\nonumber\\
&&
(G_{12,t/2}^{p_1p_1^\prime})_{ij} \,
(G_{21,-t/2}^{p_2 p_2^\prime})_{jk},\label{rg_L}
\end{eqnarray}
with $\delta_\omega=\delta(\Lambda-|\omega|)$. $(G)_{ij}=\langle i|G|j\rangle$ 
denotes the matrix element with respect to the eigenvectors of $L_{\text{D}}$,
\begin{equation}
\label{eigenvectors}
L_\text{D}|j\rangle = \lambda_j |j\rangle, \quad\quad \lambda_j=h_j-i\Gamma_j.
\end{equation}
$h$ and $\Gamma>0$ describe 
dot excitations and decay rates, respectively. Since $\Gamma$ leads to exponential damping
between the vertices, the time integrals can be cut off 
by $\Gamma^{-1}$, thereby neglecting only small perturbative corrections for
energy scales below $\Gamma$. Note, however, that $L_{\text{D}}$ has a unique
eigenvector $|0\rangle$ with zero eigenvalue since the system is approaching a stationary state.
\cite{com3} Therefore, the contribution from this eigenvector does not lead to
exponential damping but it will be shown below that it does not contribute to the
RG flow in leading order. For later purpose, we note that 
the ``ket'' form $\langle ss'|0\rangle$ depends on the specific problem
under consideration, but the ``bra'' form $\langle 0|ss'\rangle$
is unique and is given by
\begin{equation}
\label{zero_eigenvector}
\langle 0|ss'\rangle = \frac1{\sqrt{Z}}\delta_{ss'},
\end{equation}
where $Z$ is the number of many-particle states considered on the dot.
This property follows directly from Eq.~(\ref{LG_property}).

Using Eqs.~(\ref{G_intpic}) and~\eqref{eigenvectors}, we obtain the following expression for
the time integral in Eq.~(\ref{rg_G}) [$\pm$ corresponds to the two
terms on the right hand side (rhs)]:
\begin{multline}
i\int_0^{\Gamma_j^{-1}}\hspace{-0.4cm}dt \, e^{\mp i(\omega_2+x_\pm)t} 
e^{-(\Gamma_j-\Gamma_{ik})t}
\\
=\,\frac{1-e^{\mp i(\omega_2+x_\pm\mp i(\Gamma_j-\Gamma_{ik}))/\Gamma_j}}
{\pm(\omega_2+x_\pm)-i(\Gamma_j-\Gamma_{ik})},\label{time_integral}
\end{multline}
with $|\omega_2|=\Lambda$, $\lambda_{ik}=(\lambda_i+\lambda_k)/2=h_{ik}-i\Gamma_{ik}$, 
$x_\pm=\mu_{\alpha_2}-\mu_{\alpha_1\alpha_1^\prime}-
\omega_{11'}\pm(h_j-h_{ik})$, 
$\mu_{\alpha_1\alpha_1^\prime}=(\mu_{\alpha_1}+\mu_{\alpha_1^\prime})/2$, and
$\omega_{11'}=(\omega_1+\omega_1^\prime)/2$.
This provides a cutoff at $\Lambda_\pm=\text{max}(|x_\pm|,\Gamma_j,|\Gamma_j-\Gamma_{ik}|)$,
containing frequencies, voltages, dot excitation energies, and decay rates.
Above the cutoffs, we obtain $\pm \text{sign}(\omega_2)/\Lambda$
for Eq.~(\ref{time_integral}). Therefore,
we can replace $-\text{sign}(\omega_2)pf^p(\omega_2)$ by 
$\text{sign}(\omega_2)p[1/2-f^p(\omega_2)]=1/2-f(\Lambda)\approx 
(1/2)\theta_T$ in Eq.~(\ref{rg_G}), providing the cutoff 
set by temperature. This gives 
\begin{eqnarray}
(dG^{p_1p_1^\prime}_{11'}/dl)_{ik}\,=\,
-\frac12\theta_T\delta_{\omega_2}&&\nonumber\\
\nonumber
&&\hspace{-2.5cm}
\times\left\{\theta_{\Lambda_+}(G^{p_1p_2}_{12})_{ij}
(G^{p_2^\prime p_1^\prime}_{21'})_{jk}\,\right.\\
\label{rg_G_cutoff}
&&\hspace{-2cm}\left.
-\theta_{\Lambda_-}(G^{p_2^\prime p_1^\prime}_{21'})_{ij}
(G^{p_1 p_2}_{12})_{jk}\right\},
\end{eqnarray}
where $l=\ln(\Lambda_0/\Lambda)$ denotes the flow paramter.
First, we get from this equation the central result that decay rates always lead
to a cutoff of the RG flow. If all $i,j,k$ correspond to the 
eigenvector with zero eigenvalue, we get $\Lambda_+=\Lambda_-$ and
the two terms on the rhs of Eq.~(\ref{rg_G_cutoff}) cancel. If at least
one eigenvector has nonzero eigenvalue, we obtain a cutoff either from
$\Gamma_j$ or $|\Gamma_j-\Gamma_{ik}|$. Neglecting the irrelevant difference
between the various decay rates, we replace them in the following 
by an overall scale $\Gamma$. Second, above all cutoff scales, the RG 
equation preserves the initial form of the vertex given by Eqs.~(\ref{G_vertex})
and~(\ref{G_initial}). Below, we show that 
a similiar analysis leads to $L_{\text{D}}=[H_{\text{D}},\cdot]$ in
leading order, with $H_{\text{D}}=\sum_s E_s|s\rangle\langle s|$ denoting the 
renormalized dot Hamiltonian [see Eq.~(\ref{H_RG})]. Inserting these forms
in Eq.~(\ref{rg_G_cutoff}), we can project the RG equation for the vertex on 
one part of the Keldysh contour and we obtain the final
result

\begin{figure}
\includegraphics[scale=1]{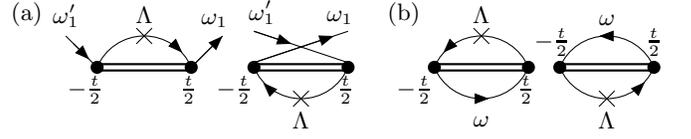}
\caption{The RG diagrams determining the renormalization of (a) $G$ and (b) $L_{\text{D}}$.
The cross indicates differentiation of the lead contractions with respect to $\Lambda$. 
Time-ordering is defined with respect to the middle time.}
\label{rg_diagrams}
\end{figure}

\begin{eqnarray}
(dg_{11'}/dl)_{ss'}&=&-\frac12\delta_{\omega_2}
\nonumber\\
&&\hspace{-2cm}\times\left\{
\theta_{\text{max}(T,|x_+|,\Gamma)}\,
(g_{12})_{s\bar{s}}(g_{21'})_{\bar{s}s'}\right.
\nonumber\\
&&\hspace{-1cm}\left.
-\,\theta_{\text{max}(T,|x_-|,\Gamma)}\,
(g_{21'})_{s\bar{s}}(g_{12})_{\bar{s}s'}
\right\}
\label{g_RG},
\end{eqnarray}
with $x_\pm=\mu_{\alpha_2}-\mu_{\alpha_1\alpha_1^\prime}-
\omega_{11'}\pm(E_{\bar{s}}-E_{ss'})$ and
$E_{ss'}=(E_s+E_{s'})/2$. If the frequency dependence is irrelevant,
we get $g_{\mu\mu'}(\omega_1,\omega_1^\prime)=g_{\mu\mu'}(\omega_{11'})$.
This equation is a generalization of the RG equation of Ref.~\onlinecite{rosch_etal} 
to an arbitrary QD in the CB 
regime with (possibly) frequency dependent density of states in the leads,
including the microscopically derived cutoff scales
from decay processes.

We now turn to the leading order analysis for the RG equation (\ref{rg_L}) of
the dot Liouvillian. We first insert the leading order form (\ref{G_vertex})
for the vertex and get
\begin{multline}
\left(\frac{dL_{\text{D}}}{d\Lambda}\right)_{ik}=
i\int_0^{\Gamma_j^{-1}} \hspace{-0.4cm} dt 
\,\frac{d}{d\Lambda}(\theta_{\omega_1}\theta_{\omega_2})
\, f^{p^\prime}_{\omega_1}\,f^{-p^\prime}_{\omega_2}
\\
(G_{12,t/2}^{pp})_{ij} \,
(G_{21,-t/2}^{p^\prime p^\prime})_{jk}.\label{rg_L_leading}
\end{multline}
Using $(d/d\Lambda)\theta_{\omega_1}\theta_{\omega_2}=
\theta_{\omega_1}\delta_{\omega_2}+\delta_{\omega_1}\theta_{\omega_2}$
and interchanging $1\leftrightarrow 2$ in the second term, we get
\begin{eqnarray}
\left(\frac{dL_{\text{D}}}{d\Lambda}\right)_{ik} &=&
i\int_0^{\Gamma_j^{-1}} \hspace{-0.4cm} dt \,
\delta_{\omega_2}\theta_{\omega_1}\nonumber\\
&&\hspace{-1.8cm}\times\left\{f^{p^\prime}_{\omega_1}
f^{-p^\prime}_{\omega_2}\, 
(G^{pp}_{12,t/2})_{ij}\,(G^{p^\prime p^\prime}_{21,-t/2})_{jk}
\,+\right.\nonumber\\
&&\hspace{-1cm}\left. +\,f^{-p}_{\omega_1} f^{p}_{\omega_2}\,
(G^{p^\prime p^\prime}_{21,t/2})_{ij}\,(G^{pp}_{12,-t/2})_{jk}
\right\}.\label{rg_L_new}
\end{eqnarray}
Performing the same steps as for the derivation of Eq.~(\ref{rg_G_cutoff}), we
obtain in leading order
\begin{eqnarray}
\left(\frac{dL_{\text{D}}}{d\Lambda}\right)_{ik} =
\,\frac1{2\Lambda}\theta_T \delta_{\omega_2}\theta_{\omega_1}&&
\nonumber\\
&&\hspace{-3cm}
\times\left\{\theta_{\Lambda_+}p' f^{p'}_{\omega_1}(G^{pp}_{12})_{ij}
(G^{p'p'}_{21})_{jk}\,+\right.\nonumber\\
\label{rg_L_cutoff}
&&\hspace{-2.5cm}\left.+\,\theta_{\Lambda_-}
p f^{-p}_{\omega_1}(G^{p'p'}_{21})_{ij}(G^{pp}_{12})_{jk}\right\},
\end{eqnarray}
with $\Lambda_\pm=\text{max}(|x_\pm|,\Gamma_j,|\Gamma_j-\Gamma_{ik}|)$ and 
$x_\pm=\mu_{\alpha_2}-\mu_{\alpha_1}-\omega_1\pm(h_j-h_{ik})$. Analogous to the
conclusion drawn from Eq.~(\ref{rg_G_cutoff}), we see that decay rates will 
always lead to a cutoff of the RG flow for $L_\text{D}$. Here, the case that 
$i\equiv 0$ corresponds to the eigenvector with eigenvalue zero can be excluded,
since $\sum_p\langle 0|G^{pp}_{12}=0$ due to Eq.~(\ref{zero_eigenvector}) and
the property (\ref{LG_property}) which is conserved under the RG flow.
Second, due to
the leading order form (\ref{G_initial}) of the vertex, 
$G^{++}_{12}=g_{12}\cdot$ acts only on the upper part of the Keldysh
contour and $G^{--}_{12}=-\cdot g_{12}$ only on the lower one. 
Therefore, they commute, and above all cutoff scales (i.e., for $\Lambda\gg\Lambda_\pm$),
we obtain no contribution from $p'=-p$ to the renormalization of $L_\text{D}$.
The contribution from $p'=p$ gives the leading order form 
$L_\text{D}^{\text{rel}}=[H_\text{D},\cdot]$ with a Hermitian renormalized dot Hamiltonian $H_\text{D}$. In 
analogy to Eq.~(\ref{g_RG}), the RG equation for $H_\text{D}$ reads
\begin{eqnarray}
(dH_{\text{D}}/d\Lambda)_{ss'} &=&\frac1{2\Lambda}
\delta_{\omega_2}\theta_{\omega_1}
\nonumber\\
&&\hspace{-2cm}\times\left\{
\theta_{\text{max}(T,|x_+|,\Gamma)}f^+_{\omega_1}\,
(g_{12})_{s\bar{s}}(g_{21})_{\bar{s}s'}\right.
\nonumber\\
&&\hspace{-1cm}\left.
+\,\theta_{\text{max}(T,|x_-|,\Gamma)}f^-_{\omega_1}\,
(g_{21})_{s\bar{s}}(g_{12})_{\bar{s}s'}
\right\}
\label{H_RG},
\end{eqnarray}
with $x_\pm=\mu_{\alpha_2}-\mu_{\alpha_1}-\omega_1\pm(E_{\bar{s}}-E_{ss'})$. 
If the frequency dependence of $g_{12}$ is irrelevant, we obtain in leading
order
\begin{equation}
\label{H_RG_approx}
(dH_{\text{D}}/d\Lambda)_{ss'}=
2\theta_{\text{max}(T,|y|,\Gamma)}\,
(g_{\mu\mu'})_{s\bar{s}}(g_{\mu'\mu})_{\bar{s}s'},
\end{equation}
with $y=\mu_\alpha-\mu_{\alpha'}-E_{\bar{s}}+E_{ss'}$.

\begin{figure}
\includegraphics[scale=1]{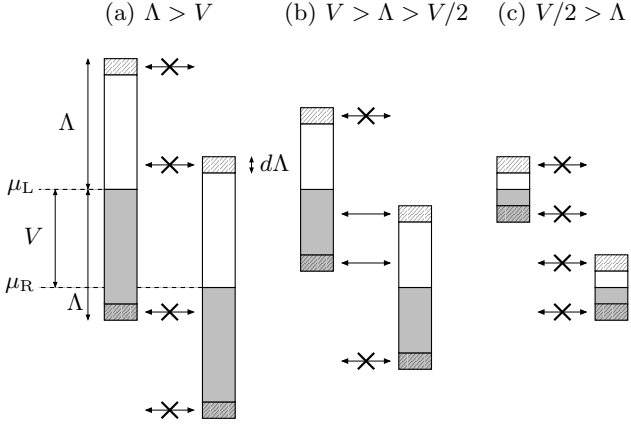}
\caption{Generation of a transition rate from state $s$ to $s'$ 
  by the RG (for simplicity, we
  consider two leads L, R with $V=\mu_\text{L}-\mu_\text{R}$ and the case
  $E_s=E_{s'}$ here). Energy-conserving transitions from one of the intervals
  $d\Lambda$ which are integrated out to the
  other lead or vice versa can only occur in situation~(b).}
\label{rate_generation}  
\end{figure}

The RG equations (\ref{g_RG}) and (\ref{H_RG}) are the central results of
this section. They provide the leading-order renormalization of the vertex 
and the dot Hamiltonian for a generic quantum dot in the Coulomb blockade regime.
Besides the full frequency dependence and the influence of temperature and
voltage, they include the influence of the renormalized dot energies and
the decay rates on the RG of the vertex. The renormalized dot energies follow
from Eq.~(\ref{H_RG}) but we still have to set up the RG equation for the decay 
rates $\Gamma_i$. They follow from Eq.~(\ref{rg_L_leading}), where we insert on the
rhs the leading order form~(\ref{G_vertex}) and~(\ref{G_initial}) for the vertex,
and the leading order form $L_\text{D}^\text{rel}=[H_\text{D},\cdot]$ for the 
dot Liouvillian, i.e., we neglect essentially the influence of the decay rates on 
themselves. As shown above, the leading order form $L_\text{D}^{\text{rel}}=[H_\text{D},\cdot]$
arises from the principal value part of the time integral and taking $p=p'$.
There are two additional contributions to $L_\text{D}$. 
The first one arises from $p=p'$ but taking the $\delta$-function
part of the time integral. This leads to a contribution of the form 
$L_{\text{D}}^{\text{br}}=\{H^{\text{br}}_{\text{D}},\cdot\}$ with an
anti-Hermitian dot Hamiltonian $H^{\text{br}}_{\text{D}}$ describing energy
broadening. The second one arises from $p=-p'$, i.e., from diagrams connecting 
the upper with the lower part of the Keldysh contour. This part is denoted by 
$L^{\text{rd}}_{\text{D}}$ and describes the physics of relaxation and dephasing. 
For $p=-p'$, only the $\delta$-function part of the time 
integral contributes and a straightforward calculation gives the results
\begin{eqnarray}
-i\left(\frac{dL^{\text{rd}}_{\text{D}}}{d\Lambda}\right)_{s_1s_1^\prime,s_2s_2^\prime} &=&
-2\pi \,\,g_{\mu\mu'}(\omega,\omega')_{s_1s_2}
g_{\mu\mu'}(\omega,\omega')_{s_1^\prime s_2^\prime}^*
\nonumber\\
&&\hspace{-1cm}
\times\frac{d}{d\Lambda}(\theta_{\omega}\theta_{\omega'})\,f^-_\omega f^+_{\omega'}
\,\,\delta(\omega-\omega'+y)
\label{rg_L_rd}
\end{eqnarray}
together with
\begin{equation}
(H^{\text{br}}_{\text{D}})_{ss'}=(-1/2)\sum_{\bar{s}}
(L^{\text{rd}}_{\text{D}})_{\bar{s}\bar{s},s's},
\label{rg_H_br}
\end{equation}
where $y=\mu_\alpha-\mu_{\alpha^\prime} + E_1 - E_2$ 
and $E_i=E_{s_is_i^\prime}$. The frequency integrals
over $\omega$ and $\omega'$ can be calculated 
analytically due to the two $\delta$-functions. If we
take the frequency dependence of the vertex in leading
order $g_{\mu\mu'}(\omega,\omega')=g_{\mu\mu'}
(\frac{\omega+\omega'}2)$, we obtain explicitly
\begin{eqnarray}
-i\left(\frac{dL^{\text{rd}}_{\text{D}}}{d\Lambda}\right)_{s_1s_1^\prime,s_2s_2^\prime} &=&
-2\pi \,\,\theta_{|y|/2}\sum_{pp'=\pm}\theta(py)f^p_\Lambda
f^{-p}_{\Lambda-|y|}\nonumber\\
\label{rg_L_rd_explicit}
&&\hspace{-3cm}
\times \, g_{\mu\mu'}(p'(\Lambda-|\frac{y}{2}|))_{s_1s_2}\,\,
g_{\mu\mu'}(p'(\Lambda-|\frac{y}{2}|))_{s_1^\prime s_2^\prime}^*.
\end{eqnarray}
Thus, for $T=0$, we see
that decay rates are only generated for $|y/2|<\Lambda<|y|$,
i.e., essentially below all cutoff scales. The reason for this interval
is a simple golden rule argument illustrated in Fig.~\ref{rate_generation}
for the case $s=s_1=s_1^\prime$, $s'=s_2=s_2^\prime$ and $E_s=E_{s'}$.
Finally, we note that including the influence of
the decay rates in the rhs of Eq.~(\ref{rg_L_rd}), one obtains 
essentially a broadening of the $\delta$-function by $\Gamma$.

A similiar analysis can be used to determine $\Sigma$, $\Sigma^\gamma$, and
$\Sigma^{\gamma\gamma'}$. However, since several boundary vertices with
different cutoff scales have
to be distinguished for the general case, we summarize here 
only those matrix elements necessary for the Kondo model [Eq.~(\ref{v_kondo})] (for
more details, see Appendix B). 
The RG equation for $\Sigma_{ss,s's'}=W_{ss'}$ is identical to the rhs of
Eq.~(\ref{rg_L_rd_explicit}) for $s\ne s'$ (note that $d\Lambda<0$, so that the 
renormalization $dW_{ss'}>0$),
\begin{eqnarray}
\nonumber
\left(\frac{dW}{d\Lambda}\right)_{ss'} &=&
-2\pi \sum_{pp'=\pm}|g_{\mu\mu'}(p'(\Lambda-|\frac{y}{2}|))_{ss'}|^2\\
\label{rg_W}
&&\hspace{0.5cm}
\times \, 
\theta_{|y|/2}\theta(py)f^p_\Lambda
f^{-p}_{\Lambda-|y|}.
\end{eqnarray}
If the nondiagonal matrix elements of the stationary distribution
are zero, the diagonal probabilities $p^{\text{st}}_s=p^{\text{st}}_{ss}$
follow from the rate equation 
\begin{equation}
\label{distribution_diagonal}
\sum_{s',\,s'\ne s}(W_{ss'}p^{\text{st}}_{s'}-W_{s's}p^{\text{st}}_{s})=0,
\end{equation}
and the stationary current can be written as 
\begin{equation}
\label{current_diagonal}
I_{\text{st}}^\gamma=\sum_{ss'}W^\gamma_{ss'}p^{\text{st}}_{s'}.
\end{equation} 
The RG for the current rate 
$\sum_s W^\gamma_{ss'}=\sum_s \Sigma^\gamma_{ss,s's'}$ is given by
the rhs of Eq.~(\ref{rg_L_rd_explicit}) but multiplied with 
$2c^\gamma_{\alpha\alpha'}=-(\delta_{\gamma\alpha}-
\delta_{\gamma\alpha'})$ and summing over $s$,
\begin{eqnarray}
\nonumber
\sum_s\left(\frac{dW^\gamma}{d\Lambda}\right)_{ss'} &=&
-2\pi \,\,\sum_{spp'} |g_{\mu\mu'}(p'(\Lambda-|\frac{y}{2}|))_{ss'}|^2\\
\label{rg_current}
&&\hspace{-0.5cm}
\times \,2\,c^\gamma_{\alpha\alpha'}\,
\theta_{|y|/2}\theta(py)f^p_\Lambda
f^{-p}_{\Lambda-|y|}.
\end{eqnarray}

To calculate the noise at finite frequency $\Omega$, we need the frequency
dependent vertices $g_{11'\Omega}$, arising from the renormalization of
the current vertex [Eq.~(\ref{I_intpic})]. They follow from Eq.~(\ref{g_RG}) by
replacing the product of the two $g$ vertices on the rhs by the
average $(1/2)(g_\Omega g + gg_\Omega)$ and shifting the cutoff $x_\sigma$
by $\mp\Omega/2$ for the two terms of this average, respectively,
\begin{eqnarray}
(dg_{11'\Omega}/dl)_{ss'}&=&-\frac14\,\delta_{\omega_2}
\nonumber\\
&&\hspace{-2.5cm}\times\left\{
\theta_{\text{max}(T,|x_+-\Omega/2|,\Gamma)}\,
(g_{12\Omega})_{s\bar{s}}(g_{21'})_{\bar{s}s'}\right.
\nonumber\\
&&\hspace{-1.5cm}\left.
-\,\theta_{\text{max}(T,|x_- -\Omega/2|,\Gamma)}\,
(g_{21'\Omega})_{s\bar{s}}(g_{12})_{\bar{s}s'}
\right.\nonumber\\
&&\hspace{-1.5cm}\left.
+\,\theta_{\text{max}(T,|x_+ +\Omega/2|,\Gamma)}\,
(g_{12})_{s\bar{s}}(g_{21'\Omega})_{\bar{s}s'}
\right.\nonumber\\
&&\hspace{-1.5cm}\left.
-\,\theta_{\text{max}(T,|x_- +\Omega/2|,\Gamma)}\,
(g_{21'})_{s\bar{s}}(g_{12\Omega})_{\bar{s}s'}
\right\}
\label{g_omega_RG}.
\end{eqnarray}
For the Kondo problem without magnetic field, it can then be shown that the 
diagonal noise follows from 
$S^{\gamma\gamma}_\Omega=\frac12\sum_{ss'}(W^{\gamma\gamma}_\Omega)_{ss'}$, 
where the RG of the noise rate
$\sum_{ss'} (W^{\gamma\gamma}_\Omega)_{ss'}=
\sum_{ss'} (\Sigma^{\gamma\gamma}_\Omega)_{ss,s's'}+(\Omega\rightarrow -\Omega)$ 
is given by the rhs of Eq.~(\ref{rg_L_rd_explicit}) but multiplied with 
$2(c^\gamma_{\alpha\alpha'})^2$, summing over $s,s'$, replacing 
$g\rightarrow g_\Omega$, $y\rightarrow y+\Omega$, and adding $\Omega\rightarrow -\Omega$,
\begin{eqnarray}
\sum_{ss'}{\left(\frac{dW^{\gamma\gamma}_\Omega}{d\Lambda}\right)}_{ss'} &=&
\label{rg_noise}\\
&&\hspace{-3.2cm}
=\,-4\pi (c^\gamma_{\alpha\alpha'})^2\,
\sum_{ss'pp'} |g_{\mu\mu'\Omega}(p'(\Lambda-|\frac{y+\Omega}{2}|))_{ss'}|^2
\nonumber \\
&&\hspace{-3cm}\times\, 
\theta_{|y+\Omega|/2}\,\theta(p(y+\Omega))\,f^p_\Lambda
\,f^{-p}_{\Lambda-|y+\Omega|}+(\Omega\rightarrow -\Omega).
\nonumber
\end{eqnarray}
The noise contribution 
from $\Sigma^\gamma_\Omega$ can be shown to be irrelevant 
without magnetic field, since $\Pi(\Omega)=i/\Omega$ in this case (see
Appendix B).

\section{Transport through single molecular magnets}
\label{sec:smm}
We now apply the formalism to transport through single molecular
magnets described by the pseudo-spin-$\frac12$ dot Hamiltonian
(\ref{H_D_kondo}) and interaction (\ref{v_kondo}), where
$g_{\mu\mu'}=(1/2)J^i_{\alpha\alpha'}
S^i\,\sigma^i_{\sigma\sigma'}$. The effective magnetic field $h$
[differing from the physical magnetic field $h_z$ used in
Eq.~(\ref{h_mol})] and the exchange interactions
are given by Eqs.~(\ref{magnetic_field})-(\ref{z_exchange_couplings}). Using the form of
the interaction in Eq.~(\ref{H_RG}), we 
obtain a constant, i.e., no contribution to
$L_{\text{D}}$. Thus, the energies, given by $E_s=sh/2$, $s=\pm$,
stay invariant. From the vertex RG equation (\ref{g_RG}),
we obtain in leading order
\begin{equation}
\frac{d}{dl}J^i_{\alpha\alpha'}(\omega)=\frac14
(\theta^i_+ + \theta^i_-)
(J^j_{\alpha\bar{\alpha}}J^k_{\bar{\alpha}\alpha'}
+J^k_{\alpha\bar{\alpha}}J^j_{\bar{\alpha}\alpha'}),
\end{equation}
where $i,j,k$ are all different, and we have defined
\begin{eqnarray}
\theta^z_\pm &=& \theta_{\text{max}(T,|x\pm h|,\Gamma)},\nonumber\\
\theta^x_\pm &=& \theta^y_\pm=\theta_{\text{max}(T,|x\pm h/2|,\Gamma)},
\end{eqnarray}
with $x=\mu_{\bar{\alpha}}-\mu_{\alpha\alpha'}-\omega$.
From Eq.~(\ref{rg_W}), we obtain for the rates
($s\ne s'$)
\begin{eqnarray}
\frac{d}{d\Lambda}W_{ss'}&=&
-(\pi/4)\sum_{ipp'\alpha\alpha'}\delta_{ss'}^i \theta_{\frac12y^{is'}_{\alpha\alpha'}}
\theta(p y^{is'}_{\alpha\alpha'})
\nonumber\\ 
&&\hspace{-1cm}\times\,
f^p_\Lambda f^{-p}_{\Lambda-|y^{is'}_{\alpha\alpha'}|}
J^i_{\alpha\alpha'}[p'(\Lambda-|y^{is'}_{\alpha\alpha'}|)]^2,
\label{rates}
\end{eqnarray}
with $\delta_{ss'}^z=\delta_{ss'}$, 
$\delta_{ss'}^{x/y}=\delta_{s,-s'}$, and
\begin{eqnarray}
y^{zs'}_{\alpha\alpha'} &=& \mu_\alpha-\mu_{\alpha'},\nonumber\\
y^{x/y,s'}_{\alpha\alpha'} &=&\mu_\alpha-\mu_{\alpha'}-s'h.
\end{eqnarray}
The decay rate and the stationary probability follow from 
$\Gamma=W_{\uparrow\downarrow}+W_{\downarrow\uparrow}$, 
$p^{\text{st}}_{\uparrow}=W_{\uparrow\downarrow}/\Gamma$, and
$p^{\text{st}}_{\downarrow}=W_{\downarrow\uparrow}/\Gamma$.\cite{com4}
Current and noise are obtained from Eq.~(\ref{rates}) as described at the
end of Sec.~\ref{sec:rg_formalism}.

\begin{figure}
\includegraphics[scale=1]{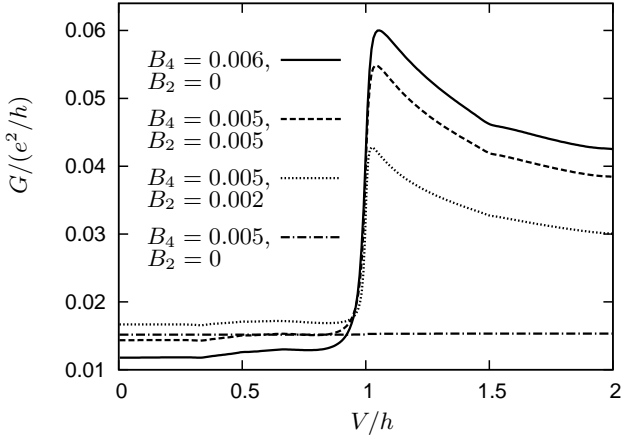}
\caption{The differential conductance as function of bias voltage
at $h_z=10^{-4}$ for different values of $B_2$ and $B_4$.
The easy-axis anisotropy of the molecule is $D=0.05$ and the coupling
to the leads $J=0.01$.
We have set $\Lambda_0=1$.}
\label{IV}  
\end{figure}

\begin{figure}
\includegraphics[width=0.45\textwidth]{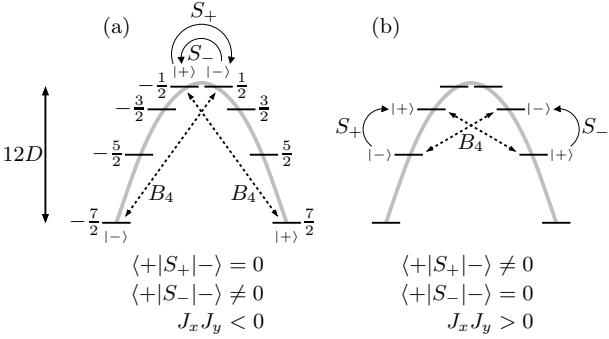}
\caption{Energy landscape and coupling of the states of a 
molecular magnet with $S=7/2$. The longitudinal anisotropy
parameter $D$ determines the height of the parabola $-DS_z^2$.
The transverse anisotropy constants $B_2$ and $B_4$ couple
every second or fourth state. For a pure $B_4$ term, this
leads to $\langle +|S_+|-\rangle$ being zero or finite
depending on the size of $B_4$ since different states form
the ground states $|\pm \rangle$.}
\label{SMM}
\end{figure} 
Fig.~\ref{IV} shows the differential conductance as function of the
voltage at finite magnetic field for different values of the $B_2$- and
$B_4$-anisotropy constants of a molecular magnet. For small 
$B_4$ and $B_2=0$, we find no Kondo effect for specific
spin values $S=3/2+2m$, $m=0,1,\dots$. In this case,
the transverse exchange couplings 
$J_{x/y}=J\langle +|S_\text{M}^+\pm S_\text{M}^-|-\rangle$
have the property $J_x J_y<0$ according to $\langle +|S_+|-\rangle=0$ [see
Fig.~\ref{SMM}(a)]. By increasing either $B_2$ or $B_4$, we get 
$\langle +|S_+|-\rangle\ne0$ [see Fig.~\ref{SMM}(b)], leading to
$J_x J_y >0$ with a quantum phase transition at $J_x J_y = 0$ to a Kondo 
effect.\cite{romeike_etal} The latter leads to
an increased conductance at $V=\pm h$ (see Fig.~\ref{IV}).

\begin{figure}
\includegraphics[scale=1]{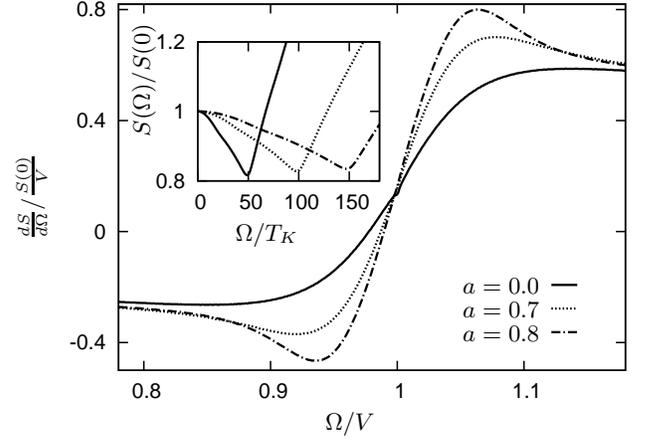}
\caption{Derivative of the noise for the isotropic Kondo model without
  magnetic field at $V=50\,T_{K,a=0}$. 
The asymmetry of the couplings is described by the asymmetry parameter $a$, where
$J_{\text{L},\text{R}}=J_0(1\pm a)$.
Inset: Noise for symmetric couplings and $V=50\,T_K$ (solid line), $V=100\,T_K$ (dotted line),
and $V=150\,T_K$ (dash-dotted line), showing a dip at $\Omega=\pm V$. }
\label{derivative}  
\end{figure}

\section{Noise for the isotropic Kondo model}
\label{sec:noise}

\begin{figure}
\includegraphics[scale=1]{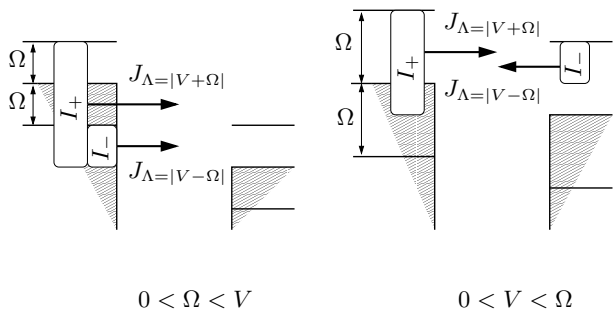}
\caption{Left: Interpretation of the finite finite frequency noise. It
  corresponds to the sum of two currents with different voltages
  $|V\pm\Omega|$ and different renormalized couplings. If both
  couplings were bare, i.e., not renormalized and therefore equal, the
  sum would be independent of $\Omega$ for $0<\Omega<V$. Right: For
  $V<\Omega$, the sum increases with $\Omega$ even for bare couplings
  because the absolute value of the currents is important.}
\label{noise_interpretation}  
\end{figure}

For $h=T=0$ and the isotropic case $J^i=J$, we obtain 
\begin{eqnarray}
\Gamma &=&\pi J^2_{\text{nd}}|_{\Lambda=V}V ,\\
I^L_\text{st} &=& \frac{3\Gamma}{4} ,\\ 
S^{LL}_\Omega &=& \frac{3\pi}{8}\sum_\pm J^2_{\text{nd},\Omega}|_{\Lambda=|V\pm\Omega|}|V\pm\Omega| ,
\end{eqnarray}
with 
\begin{equation}
\frac{d J_{\text{nd},\Omega}}{dl} = \sum_\pm\theta_{\text{max}(|V\pm\Omega)|,\Gamma)}
J_{\text{d}} J_{\text{nd}}
\end{equation}
in leading order, where 
$J_{\text{nd}}=J_{LR}=J_{RL}$ and $J_{\text{d}}=J_{LL}=J_{RR}$
for a symmetric coupling to the leads. Whereas the decay rate and the
current are cut off by the voltage, the noise
is cut off by $|V\pm\Omega|$ which can be tuned to
$0$ by setting $\Omega = \pm V$. As a result, the noise is sensitive to
the cutoff $\Gamma$ of the couplings at these points (see
Fig.~\ref{derivative}).
There is a simple interpretation of the shape of the noise: 
It can be interpreted in terms of a golden rule expression with
two superimposed currents 
$\sim J^2_{\text{nd},\Omega}|_{\Lambda=|V\pm\Omega|}|V\pm\Omega|$
with renormalized couplings (see Fig.~\ref{noise_interpretation}).
For $\Omega<V$, the sum of both currents would be independent of $\Omega$ for bare couplings, but this
balancing does not hold for renormalized couplings and a dip evolves
(see inset of Fig.~\ref{derivative}), whereas for $\Omega>V$ the noise
rises again. The effect of $\Gamma$ is prominent in the vicinity of $\Omega=
\pm V$. In Fig.~\ref{derivative}, the derivative $dS(\Omega)/d\Omega$
is shown. The shoulders around $\Omega = V$ show the logarithmic
scaling of the coupling. By tuning the asymmetry, the relaxation
rate $\Gamma$ can be tuned which results in different cutoff heights of
the shoulders, leaving the remainder of the noise untouched.

\section{Conclusion and outlook}
\label{sec:conclusion}

In this work, we have discussed a fundamental model of dissipative
quantum mechanics: a local quantum system at fixed particle number
coupled via spin or orbital exchange to several electronic reservoirs
(the generalization to bosonic reservoirs is straightforward and
goes along similiar lines). We have proposed that a microscopic derivation of
cutoff scales from decay rates should be based on a formulation in terms of 
the reduced density matrix of the local system, since the decay rates 
occur naturally by their definition,
namely, as the negative imaginary part of the eigenvalues of the kernel
determining the time evolution of the reduced density matrix. We have
shown that a complete description of decay rates in RG formalism is only 
possible if one considers the full Keldysh structure, since relaxation
and dephasing essentially
arise from diagrams connecting the upper with the lower part of the
Keldysh contour. Therefore, projecting the RG equation from the very
beginning on only one part of the Keldysh contour, one cannot obtain
a microscopic description of decay rates. Although
energy broadening terms might still lead to a cutoff of the projected
RG flow in this case, the physics of relaxation and dephasing is not included.
Therefore, our approach provides a consistent nonequilibrium
RG formulation which can identify the generation of energy broadening, 
relaxation, and dephasing at the same time, together with their influence
on the RG flow of the vertices.

Within our formalism, a particular problem arises due to the existence of
an eigenvector with zero eigenvalue (the stationary state, which is
always present and unique, at least in the absence of symmetry-breaking).
It is a nontrivial technical issue to show that this eigenvector 
does not induce a flow to strong coupling. We have achieved this for a 
generic quantum dot in the Coulomb
blockade regime by analyzing the one-loop RG equations in leading order,
including all kinds of boundary vertices determining the dot distribution,  
the current, and the noise in frequency (Laplace) space. From a pure 
physical point of view, one does not expect that the presence of a 
stationary state is correlated to the presence of a strong coupling fixed 
point, since the former is generic and the latter model-specific.
Therefore, we believe that our leading-order analysis will hold in all
orders but a general technical proof for this is still lacking.
Nevertheless, we have demonstrated within a one-loop
formulation that decay rates cut off the RG flow generically. The 
precise prefactor of the various decay rates cannot be determined
by our analytic formulation so far, since certain irrelevant 
contributions of higher-order terms have been included in the
one-loop equations, arising from the $e^{\pm iL_\text{D}t}$-factors
in the definition of the interaction picture [see, e.g., 
Eq.~(\ref{G_intpic})]. These factors are essential to provide a cutoff 
scale for the generic case, but lead also to the unwanted effect that 
certain combinations of decay rates occur which prohibit an unambigious 
determination of the correct prefactor. Also here, further 
developments of nonequilibrium RG are needed to provide generic
schemes for problems where the prefactor of the decay rates plays
an important role, e.g., for problems with several decay rates 
differing by many orders of magnitude.

The formulation of this work is useful for problems which stay in the
weak-coupling regime with decay rates of the same order of magnitude.
In this case, the different prefactors will only lead to very weak
logarithmic corrections. We have obtained the physically very natural
result that decay rates are only generated when the band width reaches
the cutoff scales, defined by voltage, temperature, frequencies, or 
dot excitations. However, considering the Kondo model, when all these 
cutoff scales are zero, the only energy scale left is the Kondo temperature 
$T_K$, and we enter the strong-coupling regime for $\Lambda < T_K$. 
An interesting issue for future research is the investigation 
of the influence of decay rates in the strong-coupling regime.
How or whether decay rates will cut off the RG flow also in this
case is an open question and has so far not been analyzed. Due to the
generic presence of decay rates, we expect them to be important
not only for weak-coupling problems. Such developments are highly desirable since
no numerical method exists so far which is capable of dealing with the 
strong-coupling limit of dissipative quantum systems in a nonequilibrium
stationary state. Benchmarks for special problems are starting to be
developed within the scattering Bethe ansatz technique,\cite{andrei_etal}
but analytical and numerical methods for generic problems are still missing.

\section*{ACKNOWLEDGMENTS}

We thank M. Keil, M. Garst, S. Jakobs, S. Kehrein, J. Paaske, A. Rosch, and 
T. Novotny for valuable discussions. This work was supported by the VW Foundation and the
Forschungszentrum J\"ulich via the virtual institute IFMIT (T.K., F.R. and H.S.).

\begin{appendix}
\section{\label{sec:rg_equations}
\bf{Derivation of THE Renormalization Group equations}}

In this appendix, we derive the RG equations (\ref{rg_G}) and (\ref{rg_L})
from Fig.~\ref{rg_diagrams}. As explained in detail in Ref.~\onlinecite{rtrg},
the $\Lambda$ dependence of the vertices $G^{p_1p_1^\prime}_{11'}$ and
the dot Liouvillian $L_{\text{D}}$ have to be defined in such a way that the
total sum of all diagrams stays invariant. This means that the derivatives
of these quantities with respect to $\Lambda$ have to cancel the 
corresponding RG diagrams of Fig.~\ref{rg_diagrams}. Using Eq.~(\ref{L_V}),
this gives for the vertex RG from Fig.~\ref{rg_diagrams}(a)
\begin{eqnarray}
-(-i)p_1^\prime\frac{dG^{p_1p_1^\prime}_{11'}}{d\Lambda} 
:J^{p_1}_{1+}J^{p_1^\prime}_{1'-}: \,\,=
(-i)^2\int_0 dt \,\,p_2 p_1^\prime \nonumber\\
\left\{G^{p_1p_2}_{12,t/2}\,G^{p_2^\prime p_1^\prime}_{2'1',-t/2}
\frac{d}{d\Lambda}
:J^{p_1}_{1+}
{J^{p_2}_{2-} J^{p_2^\prime}_{2'+}
  \begin{picture}(-20,11) 
    \put(-30,10){\line(0,1){3}} 
    \put(-30,13){\line(1,0){20}} 
    \put(-10,10){\line(0,1){3}}
  \end{picture}
  \begin{picture}(20,11) 
  \end{picture}}
J^{p_1^\prime}_{1'-}:
\right. \nonumber\\
\left. +\,G^{p_2^\prime p_1^\prime}_{2'1',t/2}\,G^{p_1p_2}_{12,-t/2}
\frac{d}{d\Lambda}
:J^{p_2^\prime}_{2'+}
{J^{p_1^\prime}_{1'-}J^{p_1}_{1+} 
  \begin{picture}(-20,11) 
    \put(-47,10){\line(0,1){3}} 
    \put(-47,13){\line(1,0){53}} 
    \put(6,10){\line(0,1){3}}
  \end{picture}
  \begin{picture}(20,11) 
  \end{picture}}
J^{p_2}_{2-}:
\right\}. \nonumber
\end{eqnarray}
Here, the contribution from the upper limit of time integration can
be shown to cancel exactly with a corresponding contribution arising
from higher order (due to certain correction terms from time ordering;
see Ref.~\onlinecite{rtrg} for further details), i.e., there is never any
divergence. Using 
\begin{equation}
\label{J_interchange}
:J^{p_1}_{1\eta}J^{p_1^\prime}_{1'\eta'}:{}=
-p_1p_1^\prime :J^{p_1^\prime}_{1'\eta'}J^{p_1}_{1\eta}:
\end{equation}
together with Eq.~(\ref{super_contraction}) and omitting the term
$:J^{p_1}_{1+}J^{p_1^\prime}_{1'-}:$ on both sides, we obtain
\begin{eqnarray}
\frac{dG^{p_1p_1^\prime}_{11'}}{d\Lambda} &=&
i\int_0 dt 
\left\{p_2^\prime \frac{d\gamma^{p_2 p_2^\prime}_{2-,2'+}}{d\Lambda}
\,G^{p_1p_2}_{12,t/2}\,G^{p_2^\prime p_1^\prime}_{2'1',-t/2}
\right. \nonumber\\
&&\hspace{0.5cm}
\left. -\,p_2 \frac{d\gamma^{p_2^\prime p_2}_{2'+,2-}}{d\Lambda}
\,G^{p_2^\prime p_1^\prime}_{2'1',t/2}\,G^{p_1p_2}_{12,-t/2}
\right\}. \nonumber
\end{eqnarray}
Inserting Eq.~(\ref{super_contraction}) for the contraction and taking
matrix elements with respect to the eigenvectors of $L_{\text{D}}$
gives the RG equation (\ref{rg_G}) for the vertex.

The RG equation for $L_{\text{D}}$ is obtained from 
Fig.~\ref{rg_diagrams}(b) and reads
\begin{eqnarray}
-(-i)\frac{dL_{\text{D}}}{d\Lambda} &=&
(-i)^2\int_0 dt \,\,p_1^\prime p_2^\prime\nonumber\\
&&G^{p_1p_1^\prime}_{11',t/2}\,G^{p_2 p_2^\prime}_{22',-t/2}
\frac{d}{d\Lambda}
:J^{p_1}_{1+}
{J^{p_1^\prime}_{1'-} J^{p_2}_{2+}
  \begin{picture}(-20,11) 
    \put(-30,10){\line(0,1){3}} 
    \put(-30,13){\line(1,0){20}} 
    \put(-10,10){\line(0,1){3}}
    \put(-47,13){\line(0,1){3}} 
    \put(-47,16){\line(1,0){53}} 
    \put(6,13){\line(0,1){3}}
  \end{picture}
  \begin{picture}(20,11) 
  \end{picture}}
J^{p_2^\prime}_{2'-}:.
\nonumber
\end{eqnarray}
Using Eqs.~(\ref{J_interchange}) and (\ref{super_contraction}), we get
\begin{eqnarray*}
\frac{dL_{\text{D}}}{d\Lambda} &=&
i\int_0 dt \,\,p_2 p_2^\prime\\
&&\frac{d}{d\Lambda}(\gamma_{1'-,2+}^{p_1^\prime p_2}
\gamma_{1+,2'-}^{p_1p_2^\prime})\,
G^{p_1p_1^\prime}_{11',t/2}\,G^{p_2 p_2^\prime}_{22',-t/2}.
\end{eqnarray*}
Again, inserting Eq.~(\ref{super_contraction}) for the contraction and taking
matrix elements with respect to the eigenvectors of $L_{\text{D}}$,
we obtain the RG equation (\ref{rg_L}) for the dot Liouvillian.

\section{\label{sec:current_noise}\bf{Current and noise}}

In order to calculate the probabilities, the current, and the noise 
from Eqs.~(\ref{distribution_omega}), (\ref{current_omega}), and
(\ref{noise_omega}), one needs RG equations for the kernels 
$\Sigma_\Omega$, $\Sigma_\Omega^{\gamma}$, and 
$\Sigma_\Omega^{\gamma\gamma'}$. The perturbation series of 
$\Sigma_\Omega$ contains terms of the following structure:
\begin{equation}
\Sigma_\Omega\rightarrow A_\Omega GG\dots GB_\Omega,
\end{equation}
with the interaction picture of the two boundary vertices $A_\Omega$ and
$B_\Omega$ defined by
\begin{eqnarray}
\label{A_intpic_anhang}
&&A^{pp'}_{11'\Omega,t}=e^{i\Omega t}e^{iEt}\,
A^{pp'}_{11'\Omega}e^{-iL_{\text{D}}t},\\ 
\label{B_intpic_anhang}
&&B^{pp'}_{11'\Omega,t}=e^{-i\Omega t}e^{iEt}\,
e^{iL_{\text{D}}t}B^{pp'}_{11'\Omega}, 
\end{eqnarray}
with $E=\omega_1-\omega_1^\prime+\mu_{\alpha_1}-\mu_{\alpha_1^\prime}$.
Initially, $A_\Omega$ and $B_\Omega$ are independent of 
$\Omega$ and given by the vertex $G$, defined in Eq.~(\ref{G_vertex}).
For the single current kernels $\Sigma_\Omega^\gamma$ and 
$(\Sigma_\Omega^\gamma)^\dagger$, we have three
different types of terms corresponding to whether the current vertex
lies at the boundaries or in the middle of a diagram,
\begin{eqnarray}
\nonumber
\Sigma_\Omega^\gamma\,&\rightarrow&\,
A^{I\gamma}_\Omega \,G\dots G\,B_\Omega\\
\nonumber
&& \,A\,G\dots G\,G_\Omega^\gamma \,G\dots G\,B_\Omega\\
&& \,A\,G\dots G\,\tilde{B}^{I\gamma}_\Omega,\\
\nonumber
{\Sigma_\Omega^\gamma}^\dagger \,&\rightarrow&\,
\,A_\Omega \,G\dots \,G B^{I\gamma}_\Omega\,\,\\
\nonumber
&& \,A_\Omega \,G\dots G\,G_{-\Omega}^\gamma \,G\dots G\,B\\
&& \tilde{A}^{I\gamma}_\Omega \,G\dots G\,B .
\end{eqnarray}
Here, $X\equiv X_{\Omega=0}$ and all current vertices 
$A^{I\gamma}_\Omega$, $\tilde{A}^{I\gamma}_\Omega$, 
$B^{I\gamma}_\Omega$, $\tilde{B}^{I\gamma}_\Omega$, and
$G^{\gamma}_\Omega$ are initially identical to the 
frequency independent current vertex $G^\gamma$, defined in 
Eq.~(\ref{I_vertex}). However, the interaction picture of all
these vertices is defined differently and, therefore, they
are no longer identical after renormalization.
With $E=\omega_1-\omega_1^\prime+\mu_{\alpha_1}-\mu_{\alpha_1^\prime}$, 
the various interaction pictures are defined by
\begin{eqnarray}
\label{AI_intpic}
&&A^{I\gamma,pp'}_{11'\Omega,t}=e^{i\Omega t}e^{iEt}\,
A^{I\gamma,pp'}_{11'\Omega}e^{-iL_{\text{D}}t},\\ 
\label{BI_intpic}
&&B^{I\gamma,pp'}_{11'\Omega,t}=e^{-i\Omega t}e^{iEt}\,
e^{iL_{\text{D}}t}B^{I\gamma,pp'}_{11'\Omega},\\ 
\label{GI_intpic}
&&G^{\gamma,pp'}_{11'\Omega,t}=e^{i\Omega t}e^{iEt}\,
e^{iL_{\text{D}}t}G^{\gamma,pp'}_{11'\Omega}e^{-iL_{\text{D}}t},\\
\label{AIt_intpic}
&&\tilde{A}^{I\gamma,pp'}_{11'\Omega,t}=e^{iEt}\,
\tilde{A}^{I\gamma,pp'}_{11'\Omega}e^{-iL_{\text{D}}t},\\ 
\label{BIt_intpic}
&&\tilde{B}^{I\gamma,pp'}_{11'\Omega,t}=e^{iEt}\,
e^{iL_{\text{D}}t}\tilde{B}^{I\gamma,pp'}_{11'\Omega}.
\end{eqnarray}
We note that the two current vertices $\tilde{A}^{I\gamma}_\Omega$
and $\tilde{B}^{I\gamma}_\Omega$ acquire only an implicit 
frequency dependence via renormalization but not an explicit
one from the interaction picture [see Eqs.~(\ref{rg_AIt}) and
(\ref{rg_BIt}) below]. 

The RG equations analogous to Eqs.~(\ref{rg_G}) and (\ref{rg_L}) follow from
Fig.~\ref{rg_diagrams} by replacing the vertex $G$ by boundary vertices
at the appropriate places,
\begin{eqnarray}
\label{rg_Sigma}
i\frac{d\Sigma_\Omega}{d\Lambda}&=& A_\Omega\times B_\Omega,\\
\label{rg_Sigma_I}
i\frac{d\Sigma^\gamma_\Omega}{d\Lambda}&=& 
A^{I\gamma}_\Omega\times B_\Omega\,+\,
A\times \tilde{B}^{I\gamma}_\Omega,\\
\label{rg_Sigma_I_dagger}
i\frac{d{\Sigma^\gamma_\Omega}^\dagger}{d\Lambda}&=& 
A_\Omega\times B^{I\gamma}_\Omega\,+\,
\tilde{A}^{I\gamma}_\Omega \times B,\\
\label{rg_Sigma_II}
i\frac{d\Sigma^{\gamma\gamma'}_\Omega}{d\Lambda}&=& 
A^{I\gamma}_\Omega\times B^{I\gamma'}_\Omega,\\
\label{rg_A}
\frac{dA_\Omega}{d\Lambda}&=& A_\Omega \cdot G,\\
\label{rg_B}
\frac{dB_\Omega}{d\Lambda}&=& G \cdot B_\Omega,\\
\label{rg_AI}
\frac{dA^{I\gamma}_\Omega}{d\Lambda}&=& 
A^{I\gamma}_\Omega \cdot G \,+\,
A\cdot G^\gamma_\Omega,\\
\label{rg_BI}
\frac{dB^{I\gamma}_\Omega}{d\Lambda}&=& 
G \cdot B^{I\gamma}_\Omega \,+\,
G^\gamma_{-\Omega}\cdot B,\\
\label{rg_AIt}
\frac{d\tilde{A}^{I\gamma}_\Omega}{d\Lambda}&=& 
\tilde{A}^{I\gamma}_\Omega \cdot G \,+\,
A_\Omega\cdot G^\gamma_{-\Omega},\\
\label{rg_BIt}
\frac{d\tilde{B}^{I\gamma}_\Omega}{d\Lambda}&=& 
G \cdot \tilde{B}^{I\gamma}_\Omega \,+\,
G^\gamma_{\Omega}\cdot B_\Omega,\\
\label{rg_GI}
\frac{dG^{\gamma}_\Omega}{d\Lambda}&=& 
G^{\gamma}_\Omega \cdot G \,+\,
G\cdot G^\gamma_\Omega,
\end{eqnarray}
where we have used the abbreviations
\begin{eqnarray}
\left((X\cdot Y)^{p_1p_1^\prime}_{11'}\right)_{ik} &\equiv&
i\frac12\theta_T
\int_0^{\Gamma_j^{-1}} \hspace{-0.4cm}dt 
\,\delta_{\omega_2}\text{sign}(\omega_2)\nonumber\\
&&\hspace{-2cm}\left\{ 
(X^{p_1p_2}_{12,t/2})_{ij}\,(Y^{p_2^\prime p_1^\prime}_{21',-t/2})_{jk}
\,+ \right.\nonumber\\
&&\hspace{-1cm}+\left.
\,(X^{p_2^\prime p_1^\prime}_{21',t/2})_{ij}\,(Y^{p_1 p_2}_{12,-t/2})_{jk}
\right\} ,\label{abbr_G}\\
(X\times Y)_{ik} &\equiv&
i\int_0^{\Gamma_j^{-1}} \hspace{-0.4cm} dt 
\,\frac{d}{d\Lambda}(\theta_{\omega_1}\theta_{\omega_2})
p_2 p_2^\prime \, f^{p_2^\prime}_{\omega_1}\,f^{-p_2}_{\omega_2}
\nonumber\\
&&
(X_{12,t/2}^{p_1p_1^\prime})_{ij} \,
(Y_{21,-t/2}^{p_2 p_2^\prime})_{jk} 
.\label{abbr_L}
\end{eqnarray}
Note that we have already used the
replacement $-pf^p_{\omega_2}\rightarrow \frac12-f_{\omega_2}
\rightarrow \frac12\theta_T\text{sign}(\omega_2)$
in Eq.~(\ref{abbr_G}). As shown in Sec.~\ref{sec:rg_formalism}, this is 
justified in leading order. We note that for the calculation of the
current and noise, given by Eqs.~(\ref{current_omega}) and 
(\ref{noise_omega}), the boundary vertices 
$A_\Omega$, $A^{I\gamma}_\Omega$, and $\tilde{A}^{I\gamma}_\Omega$
are only needed by summing over the Keldysh indices,
\begin{equation}
\label{A_no_Keldysh}
X=\sum_{pp'}X^{pp'}\quad\quad \text{for}\,X=A_\Omega,
A^{I\gamma}_\Omega, \tilde{A}^{I\gamma}_\Omega.
\end{equation}

We find that the property (\ref{LG_property}) is preserved under
RG and holds also for the
vertices $A_\Omega$ and $B_\Omega$ and for the kernel $\Sigma_\Omega$,
which by using Eq.~(\ref{zero_eigenvector}) reads
\begin{eqnarray}
\label{ABG_prop}
\sum_{pp'}\langle 0|X^{pp'}_{11'}\,&=&\,0
\quad\text{for}\quad X=A_\Omega,B_\Omega,G ,\\
\label{L_Sigma_prop}
\langle 0|X\,&=&\,0
\quad\text{for}\quad X=L_\text{D},\Sigma_\Omega.
\end{eqnarray}
From these properties, we get directly
\begin{eqnarray}
\label{A_G}
\langle 0|A^{pp'}_\Omega &=& \langle 0|G^{pp'}_\Omega ,\\
\label{AI_GI}
\langle 0|A_\Omega^{I\gamma,pp'} &=& \langle 0|G_\Omega^{\gamma,pp'} ,\\
\label{AIt_GI}
\sum_{pp'}\langle 0|\tilde{A}^{I\gamma,pp'}_\Omega 
&=& \sum_{pp'} \langle 0| G^\gamma_{\Omega =0} ,
\end{eqnarray}
since these quantities fulfill the same RG equation and have
the same initial condition.

We now show in leading order that all RG equations are cut off by 
the decay rate $\Gamma$. The proof used in Sec.~\ref{sec:rg_formalism}
to show that the vertex $G$ is cut off by $\Gamma$ can be applied in a 
similiar way 
to the RG equation (\ref{rg_GI}) for the vertex $G_\Omega^\gamma$. For
the boundary vertices, this proof does not work since the two vertices
on the rhs of the RG equations (\ref{rg_A})-(\ref{rg_BIt}) are not equal
and the interaction picture of the boundary vertices differs from the
one of $G$ due to the absence of either $e^{iL_\text{D}t}$ (for $A_\Omega$, 
$A^{I\gamma}_\Omega$, and $\tilde{A}^{I\gamma}_\Omega$) to the left or
$e^{-iL_\text{D}t}$ (for $B_\Omega$, $B^{I\gamma}_\Omega$, and 
$\tilde{B}^{I\gamma}_\Omega$) to the right. However, we can make use
of the property (\ref{ABG_prop}) to show that the eigenvector with
eigenvalue zero cannot influence the RG equations in leading order.
To show this, we consider as an example an RG equation of the form
\begin{eqnarray}
\label{rg_X_1}
\hspace{-0.5cm}
\frac{dX^{pp'}}{d\Lambda}&=& (G \cdot X)^{pp'}\nonumber\\
\label{rg_X_2}
\hspace{-0.5cm}
&=& (G^{p\bar{p}}\cdot X^{\bar{p}'p'})^{(1)}\,+\,
(G^{\bar{p}'p'}\cdot X^{p\bar{p}})^{(2)},
\end{eqnarray}
where the two terms on the rhs of (\ref{rg_X_2}) correspond
to the two terms on the rhs of Eq.~(\ref{abbr_G}), and
the interaction picture of $X$ is definded according to the
boundary vertex $B$ [for terms contributing to the boundary
vertex $A$, the analysis is even simpler, since only the form 
summed over the Keldysh indices is needed; see Eq.~(\ref{A_no_Keldysh})]. 
Defining
\begin{align*}
X&=\sum_{pp'}X^{pp'}, & X_L&=\sum_{pp'}pX^{pp'},\\
X_R&=\sum_{pp'}p'X^{pp'}, & X_{LR}&=\sum_{pp'}pp'X^{pp'},
\end{align*}
and the same for $X\rightarrow G$, we get 
\begin{eqnarray}
\label{rg_X}
\frac{dX}{d\Lambda}&=& 
(G\cdot X)^{(1)}\,+\,
(G\cdot X)^{(2)},\\
\label{rg_XL}
\frac{dX_L}{d\Lambda}&=& 
(G_L\cdot X)^{(1)}\,+\,
(G\cdot X_L)^{(2)},\\
\label{rg_XR}
\frac{dX_R}{d\Lambda}&=& 
(G\cdot X_R)^{(1)}\,+\,
(G_R\cdot X)^{(2)},\\
\label{rg_XLR}
\frac{dX_{LR}}{d\Lambda}&=& 
(G_R\cdot X_L)^{(1)}\,+\,
(G_L\cdot X_R)^{(2)}.
\end{eqnarray}
We now consider the contribution when the two eigenvectors $i=j=0$ in
Eq.~(\ref{abbr_G}) have zero eigenvalue, i.e., $\lambda_i=\lambda_j=0$ 
(note that the eigenvalue $\lambda_k$ cannot lead to any cutoff since
the factor $e^{-iL_\text{D}t}$ to the right is missing in the definition of the interaction
picture of $X$). Due to $\langle 0|G=\langle 0|X=0$, we see that this term does not
lead to any contribution in Eqs.~(\ref{rg_X})-(\ref{rg_XR}). Only
for (\ref{rg_XLR}) can the case $i=j=0$ contribute and no cutoff from a decay rate
occurs. However, since the vertices $G_L$, $G_R$, $X_L$, and $X_R$ are
cut off by the decay rate, this term leads only to a logarithmic correction to
$X_{LR}$ which is integrable and subleading (note that frequencies enter the
argument of the logarithm, making this term finite when ingrated over the
frequencies). Therefore, we see that we do not have to consider the eigenvalue
zero in leading order (we expect that a similiar proof holds in all orders but
this cannot be seen from one-loop RG equations).

A similiar analysis can be performed for all terms of the RG equations
(\ref{rg_A})-(\ref{rg_BIt}). With this result, we can replace 
$L_\text{D}\rightarrow L_\text{D}^{\text{rel}}=[H_\text{D},\cdot]$ on the rhs of 
Eq.~(\ref{abbr_G}) and introduce an overall cutoff factor $\theta_\Gamma$, 
\begin{eqnarray}
(X\cdot Y)^{p_1p_1^\prime}_{11'} &\equiv&
i\frac12\theta_\Gamma\theta_T \int_0 dt 
\,\delta_{\omega_2}\text{sign}(\omega_2)\nonumber\\
&&\hspace{-2cm}\left\{ 
X^{p_1p_2}_{12,t/2}\,Y^{p_2^\prime p_1^\prime}_{21',-t/2}
\,+ \right.\nonumber\\
&&\hspace{-1.5cm}+\left.
\,X^{p_2^\prime p_1^\prime}_{21',t/2}\,Y^{p_1 p_2}_{12,-t/2}
\right\}_{L_\text{D}\rightarrow L_\text{D}^{\text{rel}}} .
\label{abbr_G_leading}
\end{eqnarray}
For Eq.~(\ref{abbr_L}), we use the same but neglect the decay rates on the
rhs because this does not cause any divergence (the resulting 
$\delta$- and principal value integrals are convergent),
\begin{multline}
X\times Y \equiv
i\int_0 dt e^{-t 0^+}
\,\frac{d}{d\Lambda}(\theta_{\omega_1}\theta_{\omega_2})
p_2 p_2^\prime \, f^{p_2^\prime}_{\omega_1}\,f^{-p_2}_{\omega_2}
\\
\times\left.X_{12,t/2}^{p_1p_1^\prime}\,
Y_{21,-t/2}^{p_2 p_2^\prime}\,\,\right|_{L_\text{D}\rightarrow L_\text{D}^{\text{rel}}}
.\label{abbr_L_leading}
\end{multline}

In the next step, we show that above all cutoff scales, the current vertices
preserve their initial form (\ref{I_vertex}) together
with Eqs.~(\ref{G_vertex}) and (\ref{G_initial}). To prove this, we evaluate 
(\ref{abbr_G_leading}) for $\Lambda$ larger than all cutoff scales and get
\begin{equation}
(X\cdot Y)^{p_1p_1^\prime}_{11'} \equiv
\frac{1}{\Lambda}\left(
X^{p_1p_2}_{12}\,Y^{p_2^\prime p_1^\prime}_{21'}
\,-\,X^{p_2^\prime p_1^\prime}_{21'}\,Y^{p_1 p_2}_{12}\right).
\label{abbr_G_relevant}
\end{equation}
Inserting this form in the RG equations (\ref{rg_A})-(\ref{rg_GI}), we
find that all vertices are independent of $\Omega$ and we get $A=B=G$ and 
$A^{I\gamma}=\tilde{A}^{I\gamma}=B^{I\gamma}=\tilde{B}^{I\gamma}=G^\gamma$.
Furthermore, inserting the initial form (\ref{I_vertex}) together with
Eq.~(\ref{G_vertex}) for the vertex 
$G^{\gamma,pp'}_{11'}=-\frac12(\delta_{\alpha_1\gamma}-
\delta_{\alpha_1^\prime\gamma})p'\delta_{pp'}G^{pp}$ on the rhs of
the RG equation (\ref{rg_GI}), we find
\begin{multline*}
\frac{dG^{\gamma,p_1p_1^\prime}_{11'}}{d\Lambda}\,=\,
-\frac12\frac{1}{\Lambda}\left\{p_1
(\delta_{\alpha_1\gamma}-\delta_{\alpha_2\gamma})
[G^{p_1p_1}_{12},G^{p_1^\prime p_1^\prime}_{21'}]\right.\\
\left.\,-\,p_1^\prime
(\delta_{\alpha_2\gamma}-\delta_{\alpha_1^\prime\gamma})
[G^{p_1^\prime p_1^\prime}_{21'},G^{p_1 p_1}_{12}]\right\}.
\nonumber
\end{multline*}
Using the initial form (\ref{G_initial}) of the vertex $G^{pp}$,
we find $[G^{++},G^{--}]=0$ and we get
\begin{eqnarray}
\frac{dG^{\gamma,p_1p_1^\prime}_{11'}}{d\Lambda}\,&=&\,
-\frac12(\delta_{\alpha_1\gamma}-\delta_{\alpha_1^\prime\gamma})
p_1\delta_{p_1 p_1^\prime}\frac{1}{\Lambda}
\left[G^{p_1p_1}_{12},G^{p_1^\prime p_1^\prime}_{21'}\right]
\nonumber\\
&=&\frac{d}{d\Lambda}\left\{
-\frac12(\delta_{\alpha_1\gamma}-\delta_{\alpha_1^\prime\gamma})
p_1\delta_{p_1 p_1^\prime}G^{p_1p_1^\prime}_{11'}\right\},
\nonumber
\end{eqnarray}
where we have used the RG equation (\ref{rg_G_cutoff}) above all
cutoff scales in the last line. This shows that the initial form
(\ref{I_vertex}) of the current vertex is preserved in leading
order. Therefore, for all vertices $X\equiv G,A^I,B^I,\tilde{A}^I,
\tilde{B}^I$, we use the form
\begin{equation}
\label{current_vertex_form}
X^{\gamma,pp'}_{11'\Omega}=c^\gamma_{\alpha_1\alpha_1^\prime}
p'X^{pp'}_{11'\Omega},
\end{equation}
also below the cutoff scales, together with
\begin{equation}
\label{X_vertex_form_1}
X^{pp'}_{11'\Omega}=\delta_{pp'}X^{pp}_{11'\Omega}
\end{equation}
for all $X\equiv G,A,B,A^I,B^I,\tilde{A}^I,\tilde{B}^I$.
Inserting the form (\ref{current_vertex_form}) for the current vertices
into the RG equations (\ref{rg_AI})-(\ref{rg_GI}) and neglecting
all terms on the rhs which do not preserve this form (and become
zero above all cutoff scales), we find the same RG equations for
the $\gamma$-independent vertices 
$A^I_\Omega,B^I_\Omega,\tilde{A}^I_\Omega,\tilde{B}^I_\Omega,G_\Omega$
but with an addtional factor $\frac12$ appearing on the rhs of the
RG equations,
\begin{eqnarray}
\label{rg_AI_2}
\frac{dA^{I}_\Omega}{d\Lambda}&=& 
\frac12\left\{
A^{I}_\Omega \cdot G \,+\,
A\cdot G_\Omega\right\}
,\\
\label{rg_BI_2}
\frac{dB^{I}_\Omega}{d\Lambda}&=& 
\frac12\left\{
G \cdot B^{I}_\Omega \,+\,
G_{-\Omega}\cdot B\right\}
,\\
\label{rg_AIt_2}
\frac{d\tilde{A}^{I}_\Omega}{d\Lambda}&=& 
\frac12\left\{
\tilde{A}^{I}_\Omega \cdot G \,+\,
A_\Omega\cdot G_{-\Omega}\right\}
,\\
\label{rg_BIt_2}
\frac{d\tilde{B}^{I}_\Omega}{d\Lambda}&=& 
\frac12\left\{
G \cdot \tilde{B}^{I}_\Omega \,+\,
G_{\Omega}\cdot B_\Omega\right\}
,\\
\label{rg_GI_2}
\frac{dG_\Omega}{d\Lambda}&=& 
\frac12\left\{
G_\Omega \cdot G \,+\,
G\cdot G_\Omega\right\}
.
\end{eqnarray}

Using Eq.~(\ref{abbr_G_leading}) in the RG equations (\ref{rg_A}), (\ref{rg_B}),
and (\ref{rg_AI_2})-(\ref{rg_GI_2}), one can easily prove the following useful 
relationships between the vertices [note that the form 
$L_\text{D}^\text{rel}=[H_\text{D},\cdot]$ implies 
$L_\text{D}^\text{rel}=(L_\text{D}^\text{rel})^\dagger$]:
\begin{eqnarray}
\label{GI_property}
{G_\Omega}^\dagger \,&=&\, G_{-\Omega},\\
\label{AB_property}
{A_\Omega}^\dagger \,&=&\, B_{\Omega},\\
\label{ABI_property}
{A^{I}_\Omega}^\dagger \,&=&\, B^{I}_{\Omega},\\
\label{ABIt_property}
(\tilde{A}^{I}_\Omega)^\dagger \,&=&\, \tilde{B}^{I}_{\Omega}.
\end{eqnarray}
Furthermore, we get the following properties for the matrix
representations of the vertices:
\begin{eqnarray}
\label{trace_rule}
\sum_p\sum_s (X^{pp}_{11'\Omega})_{ss,\cdot\cdot}\,&=&\,0,\\
\label{interchange_rule}
(X^{pp}_{11'\Omega})_{s_1 s_1^\prime,s_2 s_2^\prime}\,&=&\,
-(X^{-p,-p}_{1'1,-\Omega})_{s_1^\prime s_1,s_2^\prime s_2}^*
\end{eqnarray}
for all vertices $X=G,A,B,A^I,B^I,\tilde{A}^I,\tilde{B}^I$.

The RG equation (\ref{rg_GI_2}) for the vertex $G_\Omega$ 
becomes especially simple. If one uses Eqs.~(\ref{abbr_G_leading})
and~(\ref{X_vertex_form_1}) and
$L_\text{D}^\text{rel}=[H_\text{D},\cdot]$,  
one finds that the initial form
\begin{equation}
\label{X_vertex_form_2}
G^{++}_{11'\Omega} = g_{11'\Omega}\cdot,\quad\quad
G^{--}_{11'\Omega} = -\cdot g_{11'\Omega}
\end{equation}
is preserved under renormalization. Therefore, one can project this
RG equation exactly on the upper Keldysh contour and obtain
\begin{equation}
\label{rg_gI}
\frac{dg_\Omega}{d\Lambda}= 
\frac12\left\{
g_\Omega \cdot g \,+\,
g\cdot g_\Omega\right\}
,
\end{equation}
where in analogy to Eqs.~(\ref{abbr_G_leading}) and (\ref{GI_intpic}),
we have defined
\begin{multline}
(x\cdot y)_{11'} \equiv
i\frac12\theta_\Gamma\theta_T \int_0 dt 
\,\delta_{\omega_2}\text{sign}(\omega_2)\\
\left\{ 
x_{12,t/2}\,y_{21',-t/2}
\,+ \,
x_{21',t/2}\,y_{12,-t/2}
\right\}
\label{abbr_g_leading}
\end{multline}
and
\begin{equation}
\label{gI_intpic}
g_{11'\Omega,t}=e^{i\Omega t}e^{iEt}\,
e^{iH_{\text{D}}t}g_{11'\Omega}e^{-iH_{\text{D}}t},\\
\end{equation}
with $E=\omega_1-\omega_1^\prime+\mu_{\alpha_1}-\mu_{\alpha_1^\prime}$.
Evaluating Eq.~(\ref{rg_gI}) gives Eq.~(\ref{g_omega_RG}) of
Sec.~\ref{sec:rg_formalism}. We note that this projection on the upper
Keldysh contour is not exactly possible for the boundary vertices 
since the interaction picture is defined differently. Whereas for
the vertex $G^{pp}_\Omega$ we can use
\begin{equation}
\nonumber
G^{++}_{11'\Omega,t}= g_{11'\Omega,t}\cdot,\quad\quad
G^{--}_{11'\Omega,t}= -\cdot g_{11'\Omega,t},
\end{equation}
a similiar equation does not hold for the boundary vertices.
Using Eqs.~(\ref{GI_property}) and (\ref{X_vertex_form_2}), we get
\begin{equation}
\label{g_symmetry}
{g_{11'\Omega}}^\dagger = g_{1'1,-\Omega}.
\end{equation}

Finally, we note that for not more than two reservoirs, 
the generation of double-current vertices must not be
considered, since they can be shown to be
irrelevant, i.e., they are not generated above all cutoff scales,
at least if the trace over the dot states and the sum over the
Keldysh indices are taken [which is the quantity entering the
noise formula (\ref{noise_omega})]. To prove this, we use the
leading order form (\ref{abbr_G_relevant}) above all cutoff
scales and get for 
the RG of double current vertices $G^{\gamma\gamma'}$
\begin{multline*}
\sum_{p_1 p_1^\prime}\text{Tr}_{\text{D}}
\frac{dG^{\gamma\gamma',p_1 p_1^\prime}_{11'}}{d\Lambda}\,=
\frac{1}{\Lambda}\sum_{p_1 p_2 p_1^\prime p_2^\prime}\text{Tr}_{\text{D}}\\
\left\{G^{\gamma,p_1p_2}_{12}G^{\gamma',p_2^\prime p_1^\prime}_{21'}
\,-\,G^{\gamma,p_2^\prime p_1^\prime}_{21'}G^{\gamma',p_1p_2}_{12}
\right\}.
\end{multline*}
Inserting the form (\ref{current_vertex_form}),
(\ref{X_vertex_form_1}), and (\ref{X_vertex_form_2})
for the current vertex, we get after some straightforward
manipulations
\begin{eqnarray}
\sum_{p_1 p_1^\prime}\text{Tr}_{\text{D}}
\frac{dG^{\gamma\gamma',p_1 p_1^\prime}_{11'}}{d\Lambda}\,&=&
\frac{1}{4\Lambda}
\left\{(\delta_{\alpha_1\gamma}-\delta_{\alpha_2\gamma})
(\delta_{\alpha_2\gamma'}-\delta_{\alpha_1^\prime\gamma'})
\right.\nonumber\\
&&\hspace{-3.5cm}
\left.-
(\delta_{\alpha_1\gamma'}-\delta_{\alpha_2\gamma'})
(\delta_{\alpha_2\gamma}-\delta_{\alpha_1^\prime\gamma})
\right\}
\text{Tr}_{\text{D}}
\left(g_{12}g_{21'}+g_{21'}g_{12}\right).
\nonumber
\end{eqnarray}
For two reservoirs, the rhs of this equation can easily
seen to be zero either for $\gamma=\gamma'$ or 
$\gamma\ne\gamma'$.

The complicated set of RG equations simplifies considerably
if one considers a problem where the dot distribution
is diagonal $(p_\Omega)_{ss'}=\delta_{ss'}(p_\Omega)_s$ and
where the dot eigenstates $|s\rangle$ do not renormalize (however,
the dot energies $E_s$ can renormalize). This is, e.g., the case 
for the fully anisotropic Kondo model under consideration in this work, 
given by Eqs.~(\ref{H_D_kondo})-(\ref{z_exchange_couplings}), due to
rotational invariance around
the $z$ axis. In this case, we need for the dot distribution 
(\ref{distribution_omega}), the current (\ref{current_omega}), 
and the noise (\ref{noise_omega}) only the matrix elements
$X_{ss,s's'}$ ($X=\Sigma_\Omega,\Sigma^\gamma_\Omega,
{\Sigma^\gamma_\Omega}^\dagger,\Sigma_\Omega^{\gamma\gamma'}$) 
for the kernels. As a consequence, we see from 
(\ref{rg_Sigma})-(\ref{rg_Sigma_II}) and (\ref{current_vertex_form})
that only the components $X_{ss,\cdot\cdot}$ 
($X=A_\Omega,A^I_\Omega,\tilde{A}^I_\Omega$) and $X_{\cdot\cdot,ss}$ 
($X=B_\Omega,B^I_\Omega,\tilde{B}^I_\Omega$) of the boundary vertices 
are needed. Using the matrix representation of 
$L_\text{D}^\text{rel}=[H_\text{D},\cdot]$ [see Eq.~(\ref{L_matrix})],  
\begin{equation}
\label{L_rel_matrix_diagonal}
(L_\text{D}^\text{rel})_{s_1 s_1^\prime,s_2 s_2^\prime} =
(E_{s_1}-E_{s_1^\prime})\delta_{s_1 s_2}
\delta_{s_1^\prime s_2^\prime},
\end{equation}
we find $(L_\text{D}^\text{rel})_{ss,s's'}=0$ and we get
\begin{eqnarray}
\label{A_AI_GI}
(A_\Omega)_{ss,\cdot\cdot}&=&(A^I_\Omega)_{ss,\cdot\cdot}
=(G_\Omega)_{ss,\cdot\cdot},\\
\label{B_BI_GI}
(B_\Omega)_{\cdot\cdot,ss}&=&(B^I_\Omega)_{\cdot\cdot,ss}=
(G_\Omega)_{\cdot\cdot,ss},
\end{eqnarray}
since the interaction pictures and the RG equations are the same for
the various quantities.
For the boundary vertices $\tilde{A}_\Omega$ and $\tilde{B}_\Omega$,
we get this property only after summing over the states and the
Keldysh indices, since otherwise the second term on the rhs of
Eqs.~(\ref{rg_AIt_2}) and (\ref{rg_BIt_2}) contributes and leads to
different renormalizations. Analogous to Eq.~(\ref{AIt_GI}), we get
\begin{eqnarray}
\label{AIt_GI_2}
\sum_{pp's}(\tilde{A}^{I,pp'}_\Omega)_{ss,\cdot\cdot}&=&
\sum_{pp's}(G^{pp'}_{\Omega=0})_{ss,\cdot\cdot},\\
\label{BIt_GI_2}
\sum_{pp's}(\tilde{B}^{I,pp'}_\Omega)_{\cdot\cdot,ss}&=&
\sum_{pp's}(G^{pp'}_{\Omega=0})_{\cdot\cdot,ss}.
\end{eqnarray}
Therefore, we need only the RG equation
(\ref{rg_gI}) for the vertex $g_\Omega$ and we can easily evaluate
Eqs.~(\ref{rg_Sigma})-(\ref{rg_Sigma_II}) by using Eqs.~(\ref{abbr_L_leading}),
(\ref{current_vertex_form}), (\ref{X_vertex_form_2}), and 
(\ref{g_property}),
\begin{multline}
\frac{d(\Sigma_\Omega)_{ss,s's'}}{d\Lambda}=
-i\frac{d}{d\Lambda}(\theta_\omega\theta_{\omega'})
f^-_\omega f^+_{\omega'}\\
\times\left\{
\frac{|g_{\mu\mu'\Omega}(\omega,\omega')_{ss'}|^2}{\Omega + \omega - \omega' + y +i\eta}\,
-\,(\Omega\rightarrow -\Omega)^*\right\},
\label{rg_Sigma_diagonal}
\end{multline}
for $s\ne s'$, and
\begin{eqnarray}
\sum_s \frac{d(\Sigma^\gamma_\Omega)_{ss,s's'}}{d\Lambda}&=&
-i\, 2\,c_{\alpha\alpha'}^\gamma 
\frac{d}{d\Lambda}(\theta_\omega\theta_{\omega'})
f^-_\omega f^+_{\omega'}\nonumber\\
&&\hspace{-3.5cm}\times\sum_s\left\{
\frac{|g_{\mu\mu'\Omega}(\omega,\omega')_{ss'}|^2}{
\Omega + \omega - \omega' + y +i\eta}\,
-\,(\Omega\rightarrow -\Omega)^*\right\},
\label{rg_Sigma_I_diagonal}\\
\sum_s \frac{d(\Sigma^{\gamma\gamma'}_\Omega)_{ss,s's'}}{d\Lambda}&=&
-i\, 2\,c_{\alpha\alpha'}^\gamma c_{\alpha\alpha'}^{\gamma'} 
\frac{d}{d\Lambda}(\theta_\omega\theta_{\omega'})
f^-_\omega f^+_{\omega'}\nonumber\\
&&\hspace{-3.5cm}\times\sum_s\left\{
\frac{|g_{\mu\mu'\Omega}(\omega,\omega')_{ss'}|^2}{
\Omega + \omega - \omega' + y +i\eta}\,
-\,(\Omega\rightarrow -\Omega)^*\right\},
\label{rg_Sigma_II_diagonal}\\
\sum_s \frac{d({\Sigma^\gamma_\Omega}^\dagger)_{ss,s's'}}{d\Lambda}&=&
\sum_s \frac{d(\Sigma^\gamma_{\Omega=0})_{ss,s's'}}{d\Lambda},
\label{rg_Sigma_I_dagger_diagonal}
\end{eqnarray}
with $y=\mu_\alpha - \mu_{\alpha'} + E_s - E_{s'}$.
For $\Omega=0$, Eq.~(\ref{rg_Sigma_diagonal}) leads to Eq.~(\ref{rg_W}) and 
Eq.~(\ref{rg_Sigma_I_diagonal}) to Eq.~(\ref{rg_current}) of Sec.
\ref{sec:rg_formalism}, giving the stationary dot distribution and the 
stationary current [the time-dependence of the dot distribution and
the current for an arbitrary initial
state can also be calculated from Eqs.~(\ref{distribution_omega}) and
(\ref{current_omega})]. The calculation for the noise simplifies
considerably if the above matrix elements do not depend on $s$ 
(which is the case, e.g., for the isotropic Kondo model in the
absence of a magnetic field due to spin symmetry). In this case,
we can average over $s$ and get from (\ref{rg_Sigma_I_dagger_diagonal})
and (\ref{L_Sigma_prop}) the explicit formulas
\begin{eqnarray}
({\Sigma_\Omega}^\dagger)_{ss,s's'}
&=&\frac12\sum_s({\Sigma_\Omega}^\dagger)_{ss,s's'}=0,\\
({\Sigma^\gamma_\Omega}^\dagger)_{ss,s's'}
&=&\frac12\sum_s({\Sigma^\gamma_\Omega}^\dagger)_{ss,s's'}
\nonumber\\
&=&\frac12\sum_s(\Sigma^\gamma_{\Omega=0})_{ss,s's'}.
\end{eqnarray}
Therefore, we get $(\Pi_\Omega)_{ss,s's'}=i/\Omega$ from
Eq.~(\ref{distribution_omega}), and
using Eqs.~(\ref{noise1}) and (\ref{noise_omega}), we obtain
for the diagonal noise [the nondiagonal one follows from Eq.~(\ref{noise_symmetry})]
for $\Omega\ne 0$
\begin{eqnarray}
\label{noise_diagonal}
S^{\gamma\gamma}_\Omega &=&
\frac{1}{Z}\sum_{ss'}
(\Sigma^{\gamma\gamma}_\Omega + \Sigma^{\gamma\gamma}_{-\Omega})_{ss,s's'}\\
&&\hspace{-1.2cm}
+\frac{1}{Z}\frac{i}{2\Omega}\left\{\sum_{ss'}
(\Sigma^{\gamma}_\Omega - \Sigma^{\gamma}_{-\Omega})_{ss,s's'}\right\}
\left\{\sum_{ss'}(\Sigma^\gamma_{\Omega=0})_{ss,s's'}\right\},
\nonumber
\end{eqnarray}
where we have used $p_s^\text{st}=1/Z$ with $Z$ denoting the number of
dot states. Using Eq.~(\ref{rg_Sigma_II_diagonal}), we get directly the RG equation
(\ref{rg_noise}) of Sec.~\ref{sec:rg_formalism} for 
$\sum_{ss'}(\Sigma^{\gamma\gamma}_\Omega+
\Sigma^{\gamma\gamma}_{-\Omega})_{ss,s's'}$. Concerning the
second term on the rhs of Eq.~(\ref{noise_diagonal}), we use
Eq.~(\ref{rg_Sigma_I_diagonal}) and interchange 
$\mu\leftrightarrow\mu'$, $\omega\leftrightarrow\omega'$ and 
$s\leftrightarrow s'$ in the second term on the rhs of this
equation. With the help of Eq.~(\ref{g_symmetry}), this gives the 
result
\begin{eqnarray}
\frac{d}{d\Lambda}\sum_{ss'}(\Sigma^\gamma_\Omega -
\Sigma^\gamma_{-\Omega})_{ss,s's'}&=&\nonumber\\
&&\hspace{-4cm}=-4i\,c_{\alpha\alpha'}^\gamma \frac{d}{d\Lambda}
(\theta_\omega\theta_{\omega'})(f_{\omega'}-f_\omega)
\nonumber\\
&&\hspace{-4cm}\times
\text{P}\left(\frac1{\Omega+\omega-\omega'+\mu_\alpha-\mu_{\alpha'}}\right)
\sum_{ss'}|g_{\mu\mu'\Omega}(\omega,\omega')_{ss'}|^2
.
\nonumber
\end{eqnarray}
Performing the integrals over $\omega$ and $\omega'$ by
neglecting the frequency-dependence of $g_{\mu\mu'\Omega}(\omega,\omega')$,
one finds that this term leads to a contribution of the order 
$\Gamma\times\text{O}(\frac{V}{\Lambda_0},\frac{\Omega}{\Lambda_0})$. 
Therefore, it is irrelevant and is left out within our
leading order analysis.
\end{appendix}



\begin{thebibliography}{99}
\bibitem{com3}
If $\Gamma_j=0$, one can show that the
contribution from the upper limit of integration
cancels exactly with a corresponding contribution from
higher order, i.e., there is never any divergence.
\bibitem{com4}
In the fully anisotropic Kondo model, there are three 
different decay rates with similiar scale. We have taken
the $1/T_1$-rate here.
\end{thebibliography}
\end{document}